\documentclass[runningheads,fleqn]{svmult}
\usepackage{makeidx}   
\usepackage{graphicx}  
\usepackage{subeqnar}  
\usepackage{multicol}  
\usepackage{taphys}    
\usepackage{rotating}
\makeindex             
\newcommand{\greeksym}[1]{{\usefont{U}{psy}{m}{n}#1}}
\newcommand{\umu}{\mbox{\greeksym{m}}}

\begin{document}
\title*{Dark Matter Direct Detection }
\toctitle{ Dark Matter Direct Detection using Cryogenic Detectors}
%
%
\titlerunning{Dark Matter Direct Detection}
%
\author{Gabriel Chardin}
\authorrunning{Gabriel Chardin}
%
%
\institute{DSM/DAPNIA/SPP, B\^{a}timent 141, CEA/Saclay\\
F-91191 Gif-sur-Yvette Cedex, France}

\maketitle              

\begin{abstract}
Solving the Dark Matter enigma represents one of the key objectives of contemporary physics. Recent astrophysical and cosmological measurements have unambiguously demonstrated that ordinary matter contributes to less than 5\% of the energy budget of our Universe, and that the nature of the remaining 95\% is unknown. Weakly Interacting Massive Particles (WIMPs)\index{WIMP} represent the best motivated candidate to fill the Dark Matter gap, and direct detection\index{direct detection} Dark Matter experiments have recently reached sensitivities allowing them to sample a first part of supersymmetric\index{supersymmetry} models compatible with accelerator constraints.

Three cryogenic experiments currently provide the best sensitivity, by nearly one order of magnitude, to WIMP interactions. With systematic uncertainties far less severe than other present techniques, the next generation of cryogenic experiments promises two orders of magnitude increase in sensitivity over the next few years. The present results, perspectives and experimental strategies of the main direct detection experiments are presented. Challenges met by future ton-scale cryogenic experiments in deep underground sites, aiming at testing most of the SUSY\index{SUSY} parameter space, are critically discussed.
\end{abstract}

The Dark Matter enigma was first formulated by Fritz Zwicky~\cite{zwicky} in the early thirties. From observations of the velocity dispersion of eight galaxies in the Coma cluster, Zwicky found that the velocities of these galaxies were far too large to allow them to remain trapped in the cluster's gravitational well. Converting the observed velocity dispersion in terms of mass deficit, Zwicky concluded that visible stars only contribute 0.5\% of the total mass present in the Coma cluster. Today, although the reevaluations of the distance scale and the Hubble parameter lead to a revised estimate of the missing mass factor, Zwicky's observations appear as one of the most astonishing and profound insights of modern physics.

\section{ Motivations for non-baryonic Dark Matter }

Determining the precise nature of Dark Matter is one of the main
open questions of contemporary physics. Its resolution will
probably entail a major Copernician revolution since baryonic\index{baryonic}
matter, which is constituting our environment, 
represents only a small fraction of the total energy content in
the Universe. But despite impressive experimental progress and effort over the last ten years~\cite{ramachers,bergstrom,morales}, the precise nature of this Dark Matter
has not yet been uncovered.

\subsection{Searching for baryonic Dark Matter}

Over the last few years, several candidates have been proposed to
solve the Dark Matter enigma. These include anomalous gravity~\cite{milgrom},
Massive Compact Halo Objects (MACHOs\index{MACHOs})~\cite{macho93,eros93}, massive neutrinos\index{neutrino}~\cite{fukuda}, axions~\cite{sikivie83} and atomic hydrogen~\cite{combes}.
The observation in 1998 of an accelerating universe~\cite{scp98,stromlo98}
has created a considerable surprise,
leading to a new incarnation of the concept of ether, called quintessence
\cite{quintessence}. With this variety of candidates, it would seem
presumptuous to believe that the solution is close at hand.

Still, the precision in the determination of the cosmological
parameters has recently dramatically increased. In particular, the value of the Hubble expansion parameter appears now fixed at a value 
of $h_{0} \sim 71 \pm 4$ km/s/Mpc~\cite{tegmark}. Also, the non-zero
cosmological constant\index{cosmological constant} term has considerably reduced the age problem of the universe, now estimated to be $\approx 13.7 \pm 0.2$ Gyr~\cite{tegmark}.

\begin{figure}[hbtp]
\includegraphics[width=.6\textwidth]{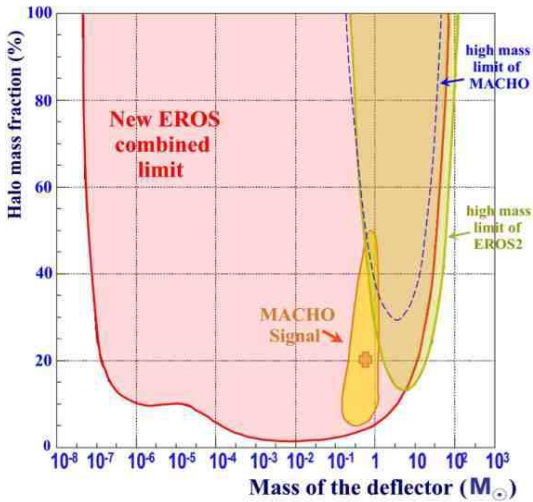}
\caption{ Exclusion diagram ($95 \%$ C.L.) on the halo mass fraction in the form of compact objects of mass M, for the standard halo model (4«10$^{11}$ Solar masses inside 50 kpc), from all LMC and SMC EROS data. The thick line is the combined limit from all EROS sub-experiments. The dotted line indicates the high mass limit of the MACHO experiment, while the enclosed region, marked by an arrow, represents the signal reported by the MACHO experiment, and excluded by EROS. For MACHOs with mass between 10$^{-6}$ and 3 solar masses, the EROS result shows that the MACHO contribution to the galactic halo is less than $10\%$
(after Ref.\protect~\cite{eros00})}	\label{EROSfinal}
\end{figure}

An impressive amount of data has been collected by
studying the rotation curves of galaxies which, apart from a few
exotic galaxies, always seem to require halos extending much
beyond the extent of luminous matter. Already at this galactic
scale, typically more than 80 \% of the matter is dark and, despite
extensive efforts, no conventional counterpart has yet been
found. For example, after the initial excitement created by the
observation of a small set of compact halo objects, detected by
their microlensing effects on very large samples of stars~\cite{macho93,eros93},
more precise measurements~\cite{macho00,eros00,eros03} over a period of nearly ten years have revealed that these MACHOs represented less than $\sim$15\% of a standard halo composed of objects with a mass between 2 $\times 10^{-7}$ M$_{sol}$ and 1 M$_{sol}$ at the 95\% C.L.

The present
experimental situation of the MACHO searches is summarized in
Fig.~\ref{EROSfinal} where the limits reached by the EROS microlensing search~\cite{eros00,eros03}, monitoring $\sim 55$ million stars in the
Small and Large Magellanic clouds,
are now excluding most of the mass range of
MACHOs. Similarly, observations of the Hubble Deep
Field Space Telescope were initially interpreted to demonstrate
that white dwarf stars, of low luminosity and intermediate mass, could represent as much as 50 \% of the galactic Dark
Matter~\cite{ibata98}. More recent data, on the other hand, indicated that this
proportion is in fact $\leq 5\%$~\cite{goldman}. Therefore, at present, no conventional candidate has been found that could explain the Galactic Dark Matter rotation curves.

\subsection{Cosmological constraints}

The Dark Matter problem is even more glaring at large scales
where the proper motions of galaxies in clusters, the study of
velocity fields, the X-ray emission temperature in
clusters, the lensing methods and the Cosmic Microwave Background (CMB) measurements all converge to indicate that
the matter content at the supercluster scale:

$$\Omega_{m}=\rho_{m}/ \rho_{c}$$

is of the order of 0.3 when expressed in terms
of the critical density $\rho_{c}$:

$$\rho_{c}=\left( 3H_{0}^{2}/8\pi G\right)\approx 1.88\times
10^{-26}h^{2}\mbox{kg m}^{-3},$$

where $H_0$ is the Hubble parameter, $G$ is the gravitational
constant and $h = H_{0}/100 \mbox{km/Mpc/s}$ is the reduced Hubble parameter.

The considerable Dark Matter densities
observed at very large scales appear to imply that a large part
of this hidden matter must be non-baryonic. This statement is
based on the success of homogeneous nucleosynthesis~\cite{schramm} and its ability to predict the light element abundances,
and notably that of helium and deuterium.

Today, the baryonic content density, expressed in terms of a
fraction of the critical density, has narrowed to~\cite{cyburt,cuoco}:

$$\Omega_{b} h^2 =0.023\pm0.001$$

with an uncertainty on the Hubble parameter $H_{0}\approx 71$ km/s/Mpc is now of the order of 5\%.

Therefore, we are led to accept that, on the one hand, a significant part of the baryons are hidden and,
secondly, that as much as 85 \% of the matter content of the
Universe is made of non-baryonic matter. The requirement of a
large non-baryonic dark matter component appears also
substantiated by the CMB data
analysis. After initial measurements at large angular scale by the COBE satellite~\cite{hinshaw}, the BOOMERANG~\cite{boomerang}, MAXIMA~\cite{maxima}, DASI~\cite{dasi}, ARCHEOPS~\cite{archeops} and CBI~\cite{cbi} balloon and ground-based experiments have constrained the total cosmological density to be within $\sim 3\%$ of the critical density. More recently, the WMAP satellite experiment~\cite{wmap} together with the SDSS and 2dF surveys~\cite{tegmark} have further reduced the uncertainty on these cosmological parameters.

\begin{figure}
\includegraphics[width=.6\textwidth]{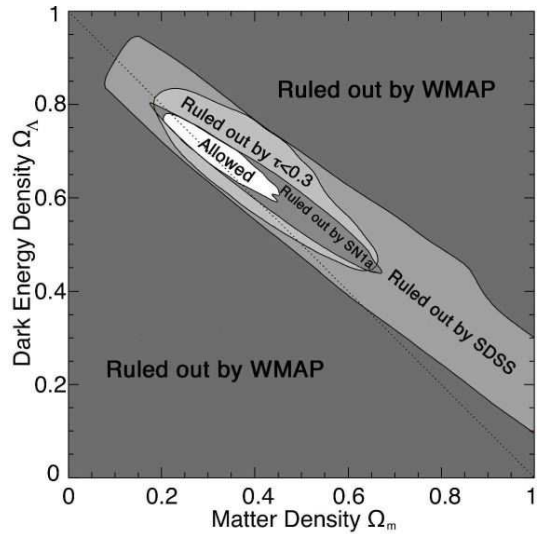}
	\caption{ Observational constraints on the matter density $\Omega_{matter}$ and on the cosmological constant density $\Omega_{\Lambda}$ ($95\%$ C.L.). Shaded dark grey region is ruled out by the Cosmological Microwave Background (CMB) WMAP satellite experiment alone. Shaded light grey region is ruled out when adding the structure constraints imposed by the Sloane Digital Sky Survey (SDSS) information. The next two regions are ruled out when adding the constraints from the reionization optical depth $\tau < 0.3$ and when including the constraints from SN-Ia supernovae, respectively. The allowed (white) region is centered on $\Omega_{matter} \sim 0.3$ and $\Omega_{\Lambda} \sim 0.7$. Models on the diagonal dotted line are flat, those below are open and those above are closed (adapted after Tegmark et al. Ref.\protect~\cite{tegmark})}
	\label{cosmological}
\end{figure}

The combined analysis of the WMAP, SDSS and 2dF data (Fig.~\ref{cosmological}) strongly favors a flat universe with a non-zero value of cosmological constant, with preferred values $\Omega_{m}\approx 0.28,\Omega_{\Lambda}\approx 0.72$

On the other hand, the recent apparition in the cosmological landscape of a non-zero cosmological constant or some other "quintessential" component has brought some uneasiness to the emerging $\Lambda$-CDM Standard Model: our Universe appears to be a strange mixture of $2/3$ of some cosmological repulsive component, $1/3$ of matter, with only a few percent of ordinary (baryonic) matter: 95\% of the universe content is unknown, while the minority matter component is composed at nearly 85\% of a weakly interacting component.

It seemed initially~\cite{initial1,initial2} that massive neutrinos could fill the Dark Matter gap. But although we have now good evidence that neutrinos are indeed massive, the sensitivity reached by present searches excludes that neutrinos can play this role: experimental constraints impose that they contribute at most a few percent to the missing mass. Generically predicted by supersymmetric (SUSY) theories, Weakly Interacting Massive Particles (WIMPs) are therefore presently considered as the best motivated candidate to solve the missing matter enigma. 

Overall, the so-called $\Lambda$-CDM "concordance" model, with a small admixture of neutrinos, reproduces nicely the available data, with a few important points on galactic central density profiles and satellite galaxies still being discussed. In addition, for the first time, direct and indirect detection\index{indirect detection} experiments are beginning to test regions of supersymmetric model parameter space compatible with cosmological and accelerator constraints.

\section{Weakly Interacting Massive Particles (WIMPs)}

\subsection{Phenomenology}

In particle physics, heavy neutrinos and supersymmetric relic
particles represent two rather natural candidates for
non-baryonic Cold Dark Matter. In fact, as noted previously and as shown by initial semiconductor experiments~\cite{initial1,initial2}, most, if not all, of the
mass range of Dirac neutrinos is already excluded experimentally. Today, the most popular candidates are given by supersymmetric
particles for two main reasons.

First, for models incorporating the conservation of a
R-parity quantum number, it is natural to expect that the lightest
supersymmetric particle will be stable. This supersymmetric particle, or
sparticle, is often supposed to be the neutralino\index{neutralino}, which is its own antiparticle and a linear combination of the photino (supersymmetric partner of the photon), the zino (partner of the Z$^0$ weak boson), the Higgsinos, and  of axinos (partners of the axion, if it exists). 

Secondly, it is a fascinating
coincidence that, for a relatively large mass range, the
cross-sections requested to produce a dark matter density $\Omega_{m} \sim 0.3$
are characteristic of electroweak interactions. In this
sense, particle physics provides a natural explanation for the
Cold Dark Matter problem. On the other
hand, the constraints on the mass of supersymmetric particles are
only phenomenological, and the LEP accelerator data only impose a
lower bound on SUSY particle masses of $\approx 35$ GeV under the MSSM\index{MSSM} hypothesis.

The phenomenology of WIMPs and in particular of the
Minimal Supersymmetric Standard Model
(MSSM) models has been described by a number
of authors~\cite{wimps}, with a rather considerable uncertainty in
the predicted event rates since more than 100 parameters remain to be
fixed to determine a specific SUSY model. Scalar\index{scalar (interaction)} and axial\index{axial (interaction)} terms
can be {\it a priori} considered for WIMP coupling to ordinary matter.
Scalar terms lead typically to an approximate $A^{2}$ dependence of
the cross-section with the number of nucleons $A$, and are therefore
usually predominant compared to axial couplings, which depend on
the spin of the target nucleus.

Interactions rates range from a few events/kg/day for the most
optimistic models, down to about 10$^{-7}$ event/kg/day. Models favored by particle theorists typically provide event rates of a few 10$^{-3}$ event/kg/day or below. Exploring a
significant part of SUSY models will therefore require an increase in sensitivity by at
least four orders of magnitude. Direct and copious
production of SUSY particles, on the other hand, is expected to be observed
at the Large Hadron Collider (LHC) around the year 2008,
but dark matter direct detection
experiments are complementary since they can test the existence
of stable weakly interacting supersymmetric particles, {\it a priori}
undetectable in an accelerator experiment.

It might seem that the existence of a cosmological constant term
contributing typically $2/3$ of the total energy content of the
universe would further increase the difficulty for
experimentalists to detect dark matter particles.
Surprisingly, the $\Omega_{\Lambda}$ term usually results in an increase in the
number of interactions with ordinary matter. This is due to the
fact that a cosmological term will have no significant
contribution to the local Dark Matter density inferred from local measurements to be $\sim$ 0.3-0.5 GeV/cm$^{3}$.
On the other hand, a lower dark
matter density at decoupling requires a higher annihilation
cross-section and, rather generally, a higher interaction
cross-section.

\begin{figure}
\includegraphics[width=.6\textwidth]{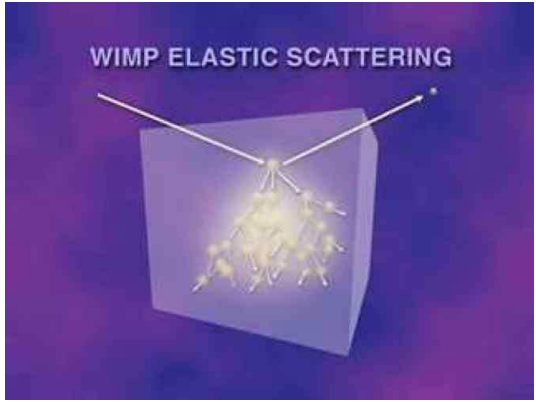}
\caption{ Schematic principle of a WIMP direct detection experiment. A WIMP scattering induces a low energy (a few keV to a few tens of keV) nuclear recoil\index{nuclear recoil}, which can be subsequently detected by the phonon\index{phonon}, charge or light signals produced in the target material.}
\label{DirectDetection}
\end{figure}

We also expect that these remnant particles will be trapped in
the Galactic gravitational potential well. Since these
particles, unlike ordinary matter, can hardly dissipate their
kinetic energy, their halo is usually
considered to be grossly spherical and non rotating (although
Sikivie has argued that infalling Dark Matter might result in
strong inhomogeneities or even caustics~\cite{caustics}). This halo would then extend
to much larger distances than the dissipating ordinary matter and
explain the rotation curves observed in galaxies. The standard
parameters used to describe the WIMP halo include its local
density in the $0.3-0.5$ GeV/cm$^3$ range, an assumption of a
Maxwellian velocity distribution with r.m.s. velocity $v_{rms} \approx 270$ km $\cdot$ s$^{-1}$ and a WIMP escape velocity from the halo $v_{esc}\approx 650$ km $\cdot$ s$^{-1}$. Using this picture of a WIMP halo, and following the seminal
paper by Drukier and Stodolsky~\cite{drukier84} on coherent
neutrino interactions, Goodman and Witten~\cite{goodman} proposed
the method of WIMP direct detection involving elastic collisions
of a WIMP on a nucleus, schematically represented on Fig.~\ref{DirectDetection}.

Except at high WIMP masses, where the momentum transfer 
requires to take into account more precisely the form factor of
the nucleus, the interaction is coherent over the whole nucleus
and the average energy in the collision can be approximated by
the expression~\cite{rich96}:
$$\langle E \rangle = m_{A}\frac{\langle v^{3} \rangle}{\langle v \rangle }\left(
\frac{m_{X}}{m_{X}+m_{A}}\right)^{2} \approx 1.6A \mbox{ keV}\left(
\frac{m_{X}}{m_{X}+m_{A}}\right)^{2}$$
where $A$ is the atomic number of the target nucleus.
It can readily be seen that for a target mass $m_A = m_X$,
this energy is approximately $0.4 \times A$ keV, giving typical energy
transfers in the few keV to a few tens of keV range for SUSY
particles with mass compatible with the constraints issued from
the LEP experiments.

\subsection{Experimental signatures}

With the previous characteristics of the WIMP halo particles, the
exponentially decreasing shape of the WIMP recoil energy spectrum
is anything but distinctive from the radioactive background,
whose energy distribution is usually also raising at low
energies. For obvious kinematical arguments, however, WIMPs give
detectable recoils almost exclusively on nuclear recoils, as
opposed to the radioactivity\index{radioactivity}, which involves mostly electron
recoils at low energies. Therefore, several methods have been
devised to distinguish as efficiently as possible nuclear from
electron recoils\index{electron recoil}. When this discrimination\index{discrimination} is achieved, the main
remaining backgrounds are due to neutrons\index{neutron} and heavy nuclear recoils associated with radon\index{radon} surface contamination.

In addition, two other signatures may be used. Firstly, in the
hypothesis of a non-rotating spherical halo, the interaction rate
is expected to be modulated by the variation in relative velocity
between the galactic halo and the Earth
in its trajectory around the Sun~\cite{annual_modulation}. The velocity
of the Earth through the Galaxy can be represented by the
following expression:

$$v_{Earth}=v_{Sun}+v_{orb} \cos \gamma \cos \left[ \omega (t-t_{0})\right],$$

where $v_{orb} \approx 30$ km s$^{-1}$ is the Earth's orbital velocity around the
Sun, the angle $\gamma \approx 60^{o}$ is the inclination of the Earth orbital plane with respect to the galactic plane, $\omega \approx 2\pi /365$ radian/day, and the phase is given by $t_{0}=\mbox{June 2}^{nd}$.

Secondly, a detector sensitive to the
recoil direction would be able to measure its diurnal modulation
due to the Earth rotation on its axis~\cite{directional}. However, at
the very low energies involved in the WIMP interactions, this
represents an ambitious objective, which has yet to be realized.

Once a WIMP signal will have been observed in a first detector
type, the interaction rates on different target materials (e.g.
Ge, Si, Xe, W, Bi...) should be compared to test the consistency of
the WIMP hypothesis, and establish the scalar, vector or axial coupling
of their interactions with ordinary matter.

A different experimental identification is provided by the
indirect detection techniques, which use the fact that WIMPs are expected to accumulate at the center of the Earth, of the Sun or
at the galactic center~\cite{sun_capture,indirect,gondolo_silk}. The WIMPs may then
annihilate in sufficient quantities for the high energy neutrino
component produced in the disintegrations to be detectable at the
surface of the Earth and identified through their directional
signature.

In the following, we will review the direct detection experiments
according to their radioactive background identification
capabilities (nuclear versus electron recoils) and
discuss the specific features of cryogenic Dark Matter experiments.

\section{ WIMP direct detection: initial results and the DAMA candidate }

For WIMPs virialized in our galaxy, typical recoil energies range from a few keV to a few tens of keV, and interaction cross-sections range from a few $10^{-6}$ to less than $10^{-11}$ picobarn. Initial direct detection experiments were unable to reach such small cross-sections and used detectors mainly dedicated to other purposes, e.g. double--beta decay search, using conventional germanium\index{germanium} detectors~\cite{initial1,initial2,heidelberg94}.

The main significant achievement of these experiments, using a set of ultrapure isotopically enriched $^{76}$Ge crystals, was the experimental demonstration that massive Dirac neutrinos could not be the solution to Dark Matter over essentially all the cosmologically relevant mass interval~\cite{initial1,initial2,heidelberg94}. Further improvements in the sensitivity of these experiments were mostly due to the passive reduction of internal $^{68}$Ge cosmogenic activation\index{activation (cosmogenic)} by deep--underground storage~\cite{heidelberg99}. Attempts to use an anti--Compton strategy resulted in the dedicated Heidelberg Dark Matter Search (HDMS), using a well--type germanium detector protecting a smaller 200 g germanium inner detector~\cite{hdms01}. Although efficient at MeV energy, this technique resulted in only a factor two gain at the lowest energies (a few keV) relevant for WIMP searches and is not yet competitive at larger WIMP mass with the previous result of the Heidelberg-Moscow experiment. The International Germanium Experiment (IGEX)~\cite{igex} is reaching a better sensitivity over most of the WIMP mass range but remains above the sensitivity level required to test the first SUSY models.

Sodium iodide NaI scintillating crystals~\cite{dama96,ukdmc96,gerbier99}, with larger target mass but much lesser energy resolutions, have also been used, notably by the DAMA, the UKDMC and the Saclay groups, to reach sensitivities of the order of 10$^{-5}$ picobarn. Despite the NaI inefficient discrimination at low energies, where the number of collected photons is small ($<\sim$ 6 per keV of electron recoil) and the scintillation\index{scintillation} time constants for electron and nuclear recoils are less separated, the DAMA experiment, using a total mass of $\sim 100$ kg of high purity NaI crystals, reported in 1998 a first indication of an annual modulation~\cite{dama9899} using a data set of $\sim$12.5 kg $\times$ year, recorded over a fraction of a year. Apart from the ELEGANT--V experiment~\cite{ejiri}, which is using NaI scintillators of total mass 730 kg as a veto for a double-beta decay experiment, the DAMA group has recently started operating the largest mass NaI detector for WIMP search using a new 250 kg NaI setup, LIBRA. Compared to ELEGANT--V, DAMA is using NaI crystals with a lower radioactive background, with differential rates at low energies of $\sim$~1-2 events/kg/keV/day down to an energy of 2 keV electron equivalent (e.e.), corresponding to $\sim 25$ keV recoil energy on iodine.

\subsection{ The DAMA annual modulation candidate }

\begin{figure}
\includegraphics[width=.6\textwidth]{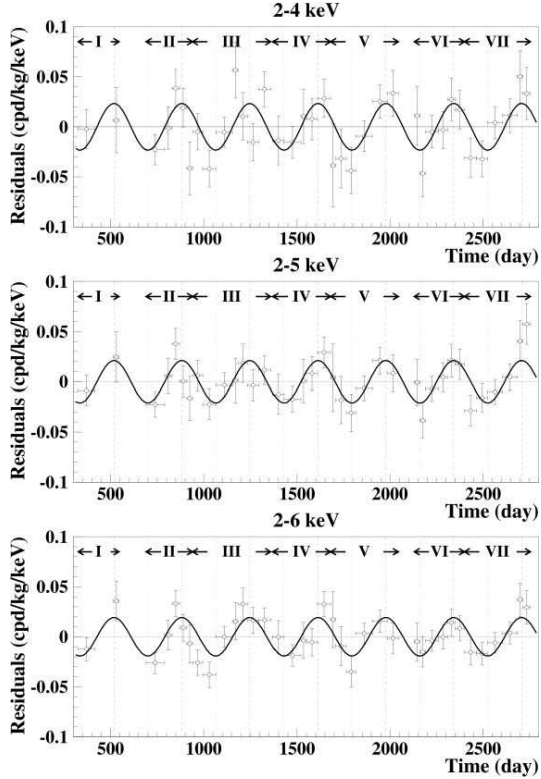}
	\caption{ Annual modulation amplitude reported by the DAMA experiment using a $\sim 10^{5}$ kg $\times$ day data sample, recorded over a period of seven years. The statistical significance of the detected modulation is $\sim$ 6.2 standard deviations (from Bernabei et al. Ref.\protect~\cite{dama03})}
	\label{dama_annual_mod}
\end{figure}

In 2000, the DAMA group published an analysis involving a 160 kg $\times$ year data sample recorded over three full annual cycles~\cite{dama00}. Recently, the group has published the analysis of three additional annual cycles~\cite{dama03} and the DAMA observation now presents a $6.2\sigma$ statistical significance, with both phase and amplitude consistent over a period of more than six years with a WIMP signature, using a 107 800 kg $\times$  day total data sample (Fig.~\ref{dama_annual_mod}). Assuming standard halo parameters~\cite{lewin_smith} and interpreted in terms of a WIMP candidate, the annual modulation would correspond to a WIMP mass $\sim (52 \pm 10)$ GeV and WIMP--nucleon cross--section $\sim (7 \pm 1) 10^{-6}$ picobarn.

EDELWEISS~\cite{edelw01,edelw02,edelw03}, in three data takings with a total exposure of 30 kg~$\times$~day was the first experiment testing and excluding a first sample of supersymmetric models. Under the assumption of spin-independent interactions, this experiment also excluded the whole DAMA region compatible with accelerator constraints without background subtraction. The DAMA group has contested this contradiction, invoking the uncertainty in the WIMP halo parameters. But Copi and Krauss~\cite{copi} have shown that, for spin-independent couplings, the contradiction remains model--independent when the relative sensitivity of both experiments is considered, unless unconventional couplings are used.

There remained the possibility that a mixture of spin-dependent\index{spin-dependent} and spin-independent\index{spin-independent} couplings could be used to reconcile the conflicting experimental results. But Kurylov and Kamionkowski~\cite{kurylov} and Savage et al.~\cite{savage} have shown that it appears impossible to reconcile the DAMA result with other negative results for all WIMP mass $> 18$ GeV.

\begin{figure}[bht]
\includegraphics[width=.6\textwidth]{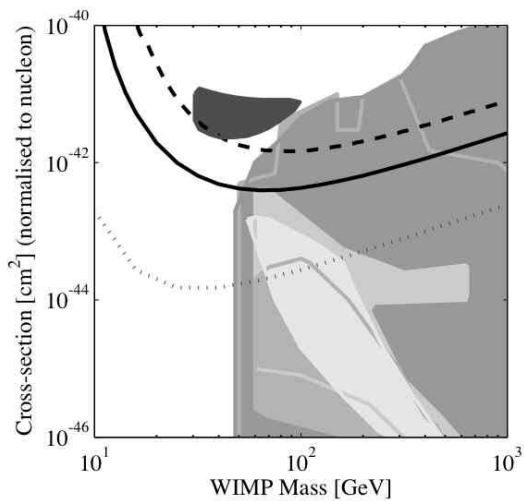}
\caption{ Experimental sensitivities of the present most sensitive WIMP direct detection experiments (after Ref.\protect~\cite{dmtools} ). The CDMS-II result (full black line), excludes the full 3-$\sigma$ zone of the DAMA signal (dark grey, upper left corner) compatible with accelerator constraints, independently of the WIMP halo model parameters. The EDELWEISS and CRESST\index{CRESST} experiments have rather similar sensitivities approximated by the black dashed line. The expected sensitivity of the next generation of cryogenic direct detection experiments is represented by the dotted line. Regions allowed by various SUSY models are represented by the light-grey regions, with WIMP-nucleon cross-sections extending down to $\sim 10^{-12}$ pbarn
\label{direct_limits}}
\end{figure}

The ZEPLIN experiment~\cite{zeplin} has used a background discrimination based on the different scintillation time constants for nuclear and electron recoils. In a 290kg~$\times$~day data sample using a 4.5 kg liquid cell, ZEPLIN announces a maximum sensitivity of $1.1 \times 10^{-6}$ picobarn. However, the electronic background rate at low energies is 50 times higher than the CDMS\index{CDMS} or EDELWEISS\index{EDELWEISS} $\gamma$-ray background rate and the claimed sensitivity requires a background subtraction at low energies by a factor $> 1000$, in a region where the nuclear/electron recoil discrimination is only statistical. Also, the energy resolution is much poorer than that of the cryogenic detectors: at 40 keV nuclear recoil energy ($\sim 6.5$ keV electron equivalent), the energy resolution is $> 100\%$, which makes it impossible to deconvolve the electron and nuclear recoil contributions. Additionally, no reliable calibrations of quenching\index{quenching factor} factor and scintillation time constants exist for nuclear recoils below $\sim$ 40 keV recoil energy, and there is a considerable discrepancy (a factor $\sim 3$) between the quenching factor measurements realized by the DAMA and by the ZEPLIN groups. These two parameters must be determined with precision before the present ZEPLIN sensitivity can be considered as established. Fig.~\ref{direct_limits} shows the constraints of the present most sensitive experiments, assuming standard halo parameters, together with the three sigma contour reported by the DAMA experiment.

\section{Cryogenic detectors: motivations}
Over the last ten years, cryogenic detectors for dark matter detection have been developed
by several groups~\cite{shutt92,lhote,bravin,meunier,coron}.
Their development was motivated by the fact that, at very low temperatures, the heat capacity approximately follows a Debye\index{Debye} law with a T$^3$ dependence and it becomes possible to
consider real calorimetric measurements down to very small energy
deposition. Energy thresholds below 1 keV of recoil energy
have already been achieved~\cite{bravin,cebrian01}. Indeed, at a
temperature of 10 mK, a 1 keV energy deposited in a 100 g
detector results in a typical temperature increase of about 1 $\umu$K,
which can be measured using conventional electronics.

\begin{figure}
\includegraphics[width=.6\textwidth]{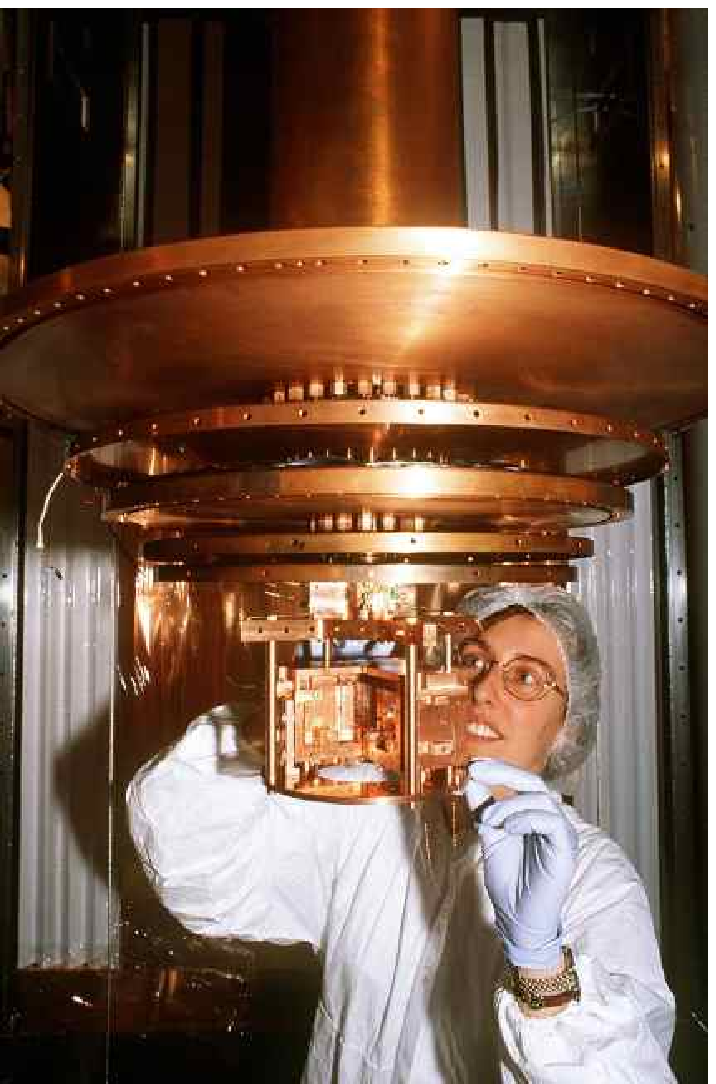}
	\caption{ Detector assembly in the CRESST low-radioactivity setup, in the Gran Sasso underground laboratory (Italy). Most of the elements close to the detectors are made of ultrapure copper, while no soldering is used. The 240 g sapphire detectors, using tungsten TES sensors, are operated at a temperature of $\sim$ 10 mK} and benefit from an energy threshold of 580 eV (99\% efficiency)
	\label{cresst_detector}
\end{figure}

In addition, the energy cost of an elementary phonon excitation is much lower ($<$~1 meV) than that of classical detectors such as semiconductors or
scintillators. Therefore, cryogenic
detectors offer the possibility of unprecedented sensitivities and energy resolutions.
The fundamental resolution of these detectors can be approximated by the
thermodynamic fluctuations in the energy of the detector :
$$\Delta E_{FWHM}\approx 2.35 \sqrt{k_{B}CT^{2}},$$
where $k_B$ is the Boltzmann constant, $C$ is the heat capacity of the
detector, and $T$ is the temperature. This resolution lies in the tens of
electron-volt range for a 100 g detector at a
temperature of 20 mK. As an explicit example, the 240 g sapphire\index{sapphire} detectors of the CRESST-I experiment, using tungsten transition edge sensors\index{transition edege sensor} (TES)\index{TES} sensors, have been operated at a temperature of $\sim$ 10 mK and displayed an energy threshold of 580 eV (99\% efficiency) (Fig.\ref{cresst_detector}).

Calorimetric measurements are realized in two rather different phonon collection modes. Although the high energy phonons
produced in a particle interaction are rapidly degraded, they benefit from
relatively long lifetimes for individual phonon energies of the
order of $10^{-4}$ eV (about 1 kelvin). These phonons can be detected
in the nearly thermal mode~\cite{cdms00,shutt92,lhote,cebrian01,alessandrello96,alessandrello97,ootani,stefano},
or when they are still out of equilibrium~\cite{bravin,young,irwin_review,cabrera}, with sub-keV energy resolutions at low energies in both cases.

But the most important impact of cryogenic detectors for Dark Matter search has been related to the background discrimination performances and the
information redundancy obtained by these detectors, with real event-by-event identification capabilities between electron recoils, associated with the gamma\index{gamma radioactivity}, beta\index{beta radioactivity} and alpha\index{alpha radioactivity} radioactive background, and nuclear recoils, observed in neutron and WIMP interactions. Table~\ref{Table1} summarizes the main characteristics of some of the main cryogenic WIMP direct detection experiments.

\begin{sidewaystable}[p]
\renewcommand{\arraystretch}{1.4}
\setlength\tabcolsep{14pt}
\begin{tabular}{llllll}
\hline\noalign{\smallskip}
Name & Location & Discrim & Target & Mass & Start \\
\noalign{\smallskip}
\hline
\noalign{\smallskip}
EDELWEISS-I & Modane (France) & Ch/Ph & Ge & 1 kg & 1996 \\
EDELWEISS-II & Modane (France) & Ch/Ph & Ge & 10-35 kg & 2005 \\
CDMS-I & Stanford (USA) & Ch/Ph & Ge/Si & 1 kg Ge, 250 g Si & 1996 \\
CDMS-II & Soudan (USA) & Ch/Ph & Ge/Si & 7 kg Ge, 1.4 kg Si & 2003 \\
CRESST-II & Gran Sasso (Italy) & L/Ph & CaWO$_4$ & 10 kg & 2003 \\
ROSEBUD & Canfranc (Spain) & L/Ph & BGO, Al$_2$O$_3$ & 0.5 kg & 2001 \\
\noalign{\smallskip}
\hline
\noalign{\smallskip}
CRESST-I & Gran Sasso (Italy) & - & Al$_2$O$_3$ & 1 kg & 1999 \\
CUORICINO & Gran Sasso (Italy) & - & TeO$_2$ & 40 kg & 2003 \\
CUORE & Gran Sasso (Italy) & - & TeO$_2$ & 760 kg & $\sim$ 2007 \\
Tokyo-DM & Kamioka (Japan) & - & LiF & 168 g & 2001 \\
ORPHEUS & Bern (Switzerland) & - & Sn & 450 g & 1995 \\
MACHe3 & Grenoble (France) & - & $^{3}$He & 0.02 g & 1998 \\
\noalign{\smallskip}
\hline
\noalign{\smallskip}
\end{tabular}\\
\smallskip
\caption{Main cryogenic WIMP direct detection experiments together with their main characteristics. The first group of experiments use background discrimination, either through simultaneous measurement of phonon and light, or of phonon and charge}
\label{Table1}
\end{sidewaystable}

\section{Charge-phonon detectors}
\begin{figure}
\includegraphics[width=.6\textwidth]{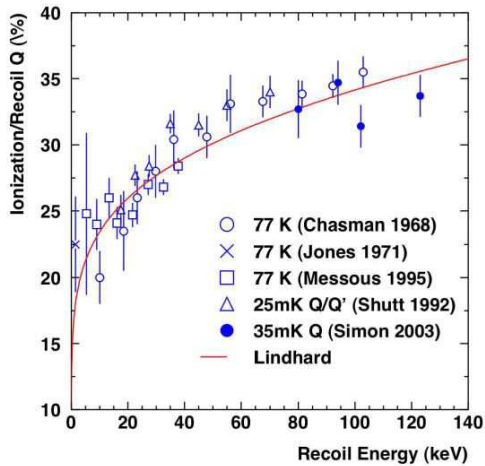}
\caption{ Variation of the ionization quenching factor in germanium as a function of recoil energy. The measurements by several groups have been realized using tagged neutron beams where the nuclear recoil energy is determined in an absolute way, by the kinematics of the reaction. The solid line is the Lindhard prediction (after Ref. \protect~\cite{sicane})
\label{ge_quenching}}
\end{figure}

Charge-phonon\index{charge-phonon} detectors provide a first discrimination method between electron recoils and nuclear recoils based on their different ionization efficiency, or quenching factor. This quenching factor is measured experimentally using neutron sources and tagged neutron beams The energy dependence of this parameter is predicted theoretically by phenomenological models, such as the Lindhard model~\cite{lindhard}. Germanium detectors, with a gap energy at low temperatures of $\sim$ 0.7 eV, require in average $\sim$ 2.9 eV of deposited electron energy to produce an electron-hole pair. For nuclear recoils, the ionization efficiency is typically a factor 3 lower than for electron recoils and depends on the deposited energy (Fig. \ref{ge_quenching}). Using cooled Field-Effect Transistors (FETs) and SQUID\index{SQUID} (the acronym of Superconducting Quantum Interference Device) electronics, these detectors present efficient discrimination performances down to recoil energies of $\sim$ 10 keV.

Following initial developments on small Si detectors~\cite{spooner91}, which demonstrated the charge-phonon discrimination, two experiments, CDMS~\cite{shutt92,clarke} and EDELWEISS~\cite{lhote,navick} have implemented this technique in Dark Matter experiments. They provide presently the best sensitivity to WIMP interactions~\cite{cdms04,edelw02}.
Two important classes of detectors can be defined: detectors with low-impedance transition edge phonon sensors (TES), sensitive to energetic out-of-equilibrium phonons, and detectors measuring the phonon signal in the thermal regime, with high-impedance sensors, either Neutron Transmuted Doped\index{NTD} (NTD) or thin films sensors\index{thin film sensor} in the Metal-Insulator Transition\index{MIT} (MIT).

\subsection{Charge-phonon detectors with MIT sensors}
Germanium charge-phonon detectors using NTDs as thermal sensors have been developed by the Berkeley group~\cite{shutt92} in the CDMS experiment, and in the EDELWEISS experiment~\cite{lhote,navick,mirabolfathi}. With respective mass of 60 and 70 grams, these prototype detectors presented excellent energy resolutions ($\sim$ 1 keV FWHM on both charge and phonon channels) and $>$ 99 \% discrimination capabilities between gamma-ray background and nuclear recoils when using neutron calibration sources (usually Am/Be or $^{252}$Cf).

\begin{figure}
\includegraphics[width=.6\textwidth]{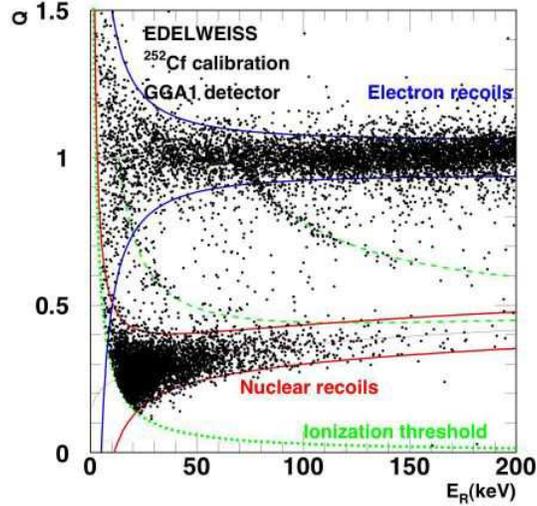}
\caption{
\label{neutron_calib}
Projection in the ($E_R$, Q) plane of the events recorded in a 320 g Ge detector of the EDELWEISS experiment during a $^{252}$Cf calibration. The thick lines represent the 90\% nuclear and electronic recoils zone ($\pm1.645\sigma$ around the neutron and gamma ionization efficiencies $<Q_n>$ and $<Q_{\gamma}>$ respectively).
The dot-dashed line is a 3~keV$_{ee}$ cut on ionisation energy.
The dashed lines show where events associated with the inelastic scattering of neutrons on $^{73}$Ge are expected (and observed)
}
\end{figure}

However, the first low-background data takings in the shallow SUF site and in the Frejus underground laboratory, under the Alps, revealed within a few days what appeared then as the major limitation of this technique: surface interactions.

\subsubsection{Solving the problem of surface events}
The problem raised by surface interactions can be understood with the help of a simple model. A particle interaction generates hot electron-hole pairs, which degrade rapidly their energy, forming a plasma of limited extension (a small fraction of 1 mm). The plasma screens the external electric field, used to collect electrons towards one electrode while holes drift towards the opposite electrode. Ramo's theorem~\cite{ramo} shows that, to a first and usually excellent approximation, the charge signal is proportional to the integral of charge drift lengths for all carriers. If the charge carriers are trapped by the wrong electrode, their contribution to the charge signal is then lost.

More precisely, the plasma screening the external collecting field is rapidly evaporated by removal of the charge carriers on the plasma outskirts, where the external field can still be partially felt. However, during this evaporation time, electron and holes suffer a random diffusion walk in the crystal. For interactions occurring within a few tens of microns from the surface, this diffusion process may bring them within reach of the metallic electrode, which will then trap charge carriers that should otherwise have drifted towards the opposite electrode.

At the low electric fields used to collect charges in a cryogenic detector (a few volts), the charge signal for a surface event is then typically 50 \% of the full collection signal since $\sim$ 50 \% of the carriers will drift and be captured by the wrong electrode. Although at energies of a few tens of keV these surface events can be efficiently separated from nuclear recoils, they become dangerously close to the signal region below $\sim$ 10 keV electron-equivalent, where a large fraction of WIMP interactions are expected.

Surface events also represent at low energies a well-known problem for classical germanium detectors~\cite{llacer}, operated at liquid nitrogen temperatures. A technique to alleviate the problem of incomplete collection was proposed by the LBL group~\cite{luke92,luke94,shutt00,shutt01}. Thin amorphous silicon\index{amorphous silicon} or germanium\index{amorphous germanium} films are sputtered on the detector surface before the deposition of metal electrodes. These amorphous films are partially conducting ($\sim$ 1 G$\Omega$ per square) and lead to a more efficient charge collection, possibly due to the modified bandgap of the amorphous layer, which can then act as a repulsive barrier for the charge carriers~\cite{shutt00}.

Attempts have also been made to increase the collecting electric field, to reduce the initial diffusion time of charge carriers when shielded by the plasma. However, it seems difficult to use collection voltages much beyond $\sim$ 10 V since the Neganov-Luke\index{Neganov-Luke effect}~\cite{neganov,luke} effect will then hide the initial phonon signal in the extra Joule heating generated by the dissipation induced by the charge drift. For high collection voltages, the phonon signal will only duplicate the charge signal information and the charge-phonon discrimination method can no longer be used.

\subsection{Charge-phonon detectors with TES sensors}
\begin{figure}
\includegraphics[width=.6\textwidth]{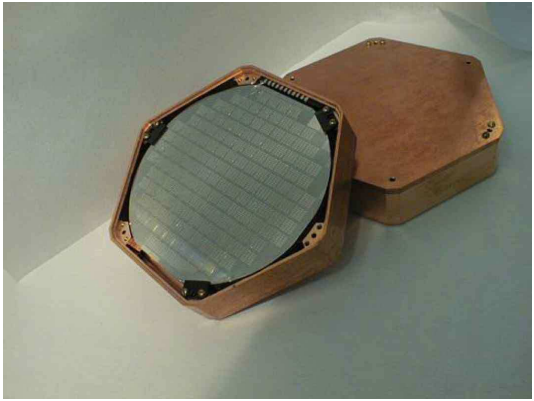}
\caption{ Photograph of a Z-sensitive ionisation-phonon (ZIP) detector of the CDMS experiment in its mount. The photolithographically-fabricated thin film on the surface is the phonon sensor, trapping non-equilibrium phonons in Al fins. The aluminum film is coupled to tungsten sensors in the superconductor-normal transition. Silicon\index{silicon} and germanium ZIPs, weighing 100 g and 250 g respectively, have been operated in the Stanford shallow undergound laboratory and are presently operated in the Soudan (Minnesota) underground laboratory
\label{zip}}
\end{figure}

The Stanford group~\cite{young,irwin_review,cabrera} has developed in the CDMS experiment a particularly elegant and integrated method using superconducting aluminum traps\index{phonon trap}~\cite{booth}, which collect phonons with individual energies exceeding the aluminum 1K  gap ($\sim$ 10$^{-4}$ eV) (Fig.~\ref{zip}). The quasiparticles\index{quasiparticle} created in the aluminum films are then partially transferred to thin ($1\umu$) tungsten sensor lines maintained by electrothermal feedback\index{electrothermal feedback}~\cite{irwin} in the normal-superconductor\index{superconductor} transition at a relatively comfortable temperature of $\sim$ 70 mK, adjusted by $^{56}$Fe implantation\index{implantation}. The aluminum 1K gap allows these detectors to be fairly insensitive to thermal phonons, and therefore to base temperature fluctuations and microphonics. In addition, the fast response of transition edge sensors, associated with a SQUID-array readout~\cite{welty}, provides a timing of the phonon signal at the few microsecond level.

Using TES sensors imposes a severe constraint on the total mass of the TES tungsten sensors due to their large heat capacity in the normal state or close to the transition. Superconducting Al fins are then used to increase the phonon collection area, thereby reducing the phonon collection time, using the phonon down-conversion and trapping technique proposed and developed by Booth~\cite{booth}. Propagation properties of trapped phonons require to use small size Al fins (380 $\umu$m $\times$ 50 $\umu$m) coupled to a large number of 1 $\umu$m  wide TES elements, implying a hierarchical structure. The design at a 1 $\umu$m precision allows a reduction of the tungsten heat capacity but requires a challenging degree of the crystal flatness after etching.

This design can be compared to that used in the CRESST experiment, where a single TES sensor, representing a small window to capture the phonon signal, induces a larger collection time. While the design of the CRESST detectors appears to lead to a better energy resolution of the phonon channel at high energies than the ZIP detectors, it does not allow position reconstruction.

These ZIP (Z-sensitive ionization-phonon) detectors benefit in their present design from four quadrant phonon sensors (Fig.~\ref{zip}). The signal amplitude repartition in these sensors is then used to obtain a millimeter position resolution of the interaction, making these detectors also interesting for real-time solar neutrino detection. The four sensors help reducing the position dependence of the phonon signal amplitude and significantly improve the energy resolution at low energies. In the 5-20 keV recoil energy interval, sub-keV resolutions are obtained on both charge and phonon channels.

In addition, these fast phonon detectors allow the identification of surface events by the relative timing between the phonon and the ionization measurements~\cite{clarke,hellmig,cdms04}. The reason for this behavior is as yet unexplained but could be possibly attributed to an anisotropic emission of ballistic phonons in the Luke process, or to the non-linear response of a phonon sensor when one of its elements becomes saturated by the energy deposited locally by a surface event.
Pragmatically, phonon sensor risetime cuts of $>$ 12 $\umu$s for Ge and $>$ 6 $\umu$s for Si provide an almost pure recoil population, at the expense of an energy-dependent fiducial cut efficiency for nuclear recoil events varying from 10-15 \% at 10 keV to 40-45 \% at 20 keV, and saturating at 50 \% above 80 keV recoil energy.

The CDMS-I experiment, at the Stanford Underground Facility (SUF), has demonstrated the discriminating properties of these ZIP detectors in a data taking of several months~\cite{cdms03}. A tower comprising four 250 gram Ge and two 100 gram ZIPs was used in this experiment. Although limited in the shallow Stanford site by the fast neutron background induced by deep-inelastic scattering of muons\index{muon (interactions)} in the surrounding rock, these detectors benefited from their very low energy threshold (down to $\sim$ 5 keV recoil energy) and therefore provided an excellent sensitivity to low mass WIMPs ($<\sim$ 40 GeV).

These detectors have then been installed in the deep site Soudan facility (Minnesota), along with a second tower comprising 4 Si and 2 Ge ZIPs. In this deep-underground site, the muon flux is attenuated by a factor $\sim 10000$ compared to that of SUF. The analysis of a first data sample of $\sim 20$ kg $\times$ day has led, using a blind analysis based on high statistics control data samples, to the best present sensitivity for all WIMP mass compatible with accelerator constraints~\cite{cdms04}.

Additional data takings by the CDMS-II experiment in Soudan using 5 towers of ZIP detectors, totalling $\sim$ 3.5 kg Ge and $\sim$ 1.5 kg Si, are expected to further improve the sensitivity to WIMP interactions by one order of magnitude.

\subsection{Thin film sensors in the metal-insulator transition}
Neutron Transmuted Doped (NTD) sensors represent an aberration for an ultra-low background experiment since these sensors are made from germanium exposed in a nuclear reactor to fast neutrons, which generate radioactive contaminants (such as $^{68}$Ge, or the relatively long-lived $^{60}$Co, and the ubiquitous tritium\index{tritium}, with its low-energy beta spectrum). The tritium can be removed from the NTD sensors by heating while the other radioactive contaminants can be identified with high precision, since they lead to electron recoils. Nevertheless, it can be expected that a small fraction of these low energy interactions will leak into the signal nuclear recoil band where the WIMP signal is expected.

Thin film sensors, with a lower mass per sensor, and using ultrapure elements, present a more controllable radioactivity. High impedance thin film sensors in the metal-insulator transition (MIT) constitute an alternative to NTD sensors, and already present excellent ($\sim$ 1 keV FWHM) energy resolution in the thermal regime~\cite{mirabolfathi}. Comparable to NTD sensors in terms of sensitivity, these film sensors may benefit in the future from large-scale production, with a manufacturing process allowing them to be deposited on a detector surface by co-evaporation under electron beam heating, without radioactive contaminants, and without the delicate manual intervention required by NTD sensors.

Thin film sensors in the MIT can be used in large surface sensors, trapping efficiently energetic phonons. Their heat capacity, for a film thickness in the 100 nm range, remains comparable to that of massive dielectric crystals in the 10 mK range. But obviously, the total volume must remain small, below typically 1 mm$^{3}$ to a few mm$^{3}$.

Another important feature of the thin MIT film sensors is, similarly to ZIP sensors, their ability to distinguish surface from volume interactions~\cite{mirabolfathi}. The different pulse shape for surface events in thin film sensors, with an additional fast component, appears to be due to an efficient trapping of out-of-equilibrium phonons, and correspondingly to a large fast athermal\index{athermal} signal for interactions close to the thin film sensor. Although allowing a rejection of surface interactions, this effect does not lead to a real 3-D position determination.

Using this technique, EDELWEISS has studied thin Nb$_{x}$Si$_{1-x}$ films which show large differences in pulseshape depending on interaction depth, with a sensitivity to surface interactions down to a depth of 0.5-1 mm~\cite{mirabolfathi}. However, like in ZIP detectors, a real determination of the z-coordinate in the detector is not yet available. The use of at least four thin film sensors covering a large fraction of the crystal surface might allow such a real 3-D position reconstruction by analysis of the relative fast vs. thermal amplitudes on each sensor.

\subsection{Charge regeneration}
For all types of charge-phonon detectors, it was found extremely important to avoid the formation of space charge in semiconductor detectors, which could reduce, or even cancel completely, the electric field inside the detector. In the non-equilibrium state of semiconductor detectors under a bias voltage at very low temperatures, a fraction of charge carriers produced in particle interactions may be trapped on acceptor and donor impurities\index{impurity} present in the crystal. If a net charge density is created in this process, there exists a critical charge density above which the electric field inside the detector is cancelled on a particular separating surface, creating a virtual electrode inside the detector. If further charge trapping occurs, the region of zero (or strongly reduced) electric field will grow, separated from the active volume of the detector by this (moving) virtual electrode.

For this reason, it was found necessary to periodically regenerate\index{charge regeneration} the detectors either by illumination by LEDs, and/or by irradiation by radioactive sources (usually $^{137}$Cs or $^{60}$Co $\gamma$-ray sources). The presence of virtual electrodes inside the detectors, delimiting a region of zero internal field, where charges created during an interaction will not drift and therefore not produce any ionization signal, can be controlled by periodic calibration runs.

Free germanium and silicon surfaces (without metal electrodes) also represent sites where charge carriers can be trapped efficiently and lead to the construction of regions with degraded fields and charge collection properties in the detector. The controlled degradation of the charge state of the CDMS and EDELWEISS detectors has been studied by the Stanford and Orsay groups, respectively~\cite{penn,broniato03,censier03}.

\begin{figure}
\includegraphics[width=.6\textwidth]{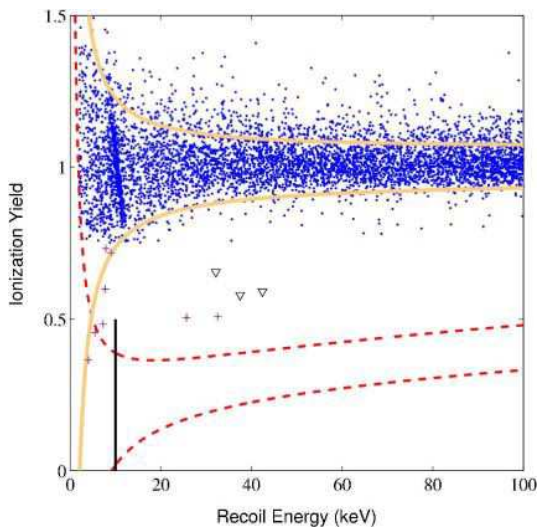}
\caption{ Scatter diagram of the ionization efficiency, normalized to electron recoils, as a function of recoil energy for all events with energy $< 100$ keV recorded by the CDMS-II experiment in the fiducial volume of 4 $\times$ 250 gram Ge detectors (after Ref.\protect\cite{cdms04}). The gamma-ray background is mostly contained within the region delimited by full lines. Nuclear recoils are expected in the region (90\% efficiency) delimited by dashed lines. No nuclear recoils are observed in a $\sim$ 20 kg $\times$ day exposure. With an effective mass 200 smaller than the DAMA NaI crystals, and an exposure 5000 times shorter, the CDMS-II exceeds by a factor $> 10$ the sensitivity of the DAMA experiment} \label{cdms04_scatplot}
\end{figure}

The use of amorphous Ge and Si underlayers, associated with periodic charge regeneration procedures, has reduced to a large extent the problem of surface events in charge-phonon detectors. Rejection factors of the $\gamma$-ray background at a $>$ 99.99 \% level have for example been obtained in the CDMS and EDELWEISS experiments in germanium detectors with both types of amorphous underlayers (Fig.~\ref{cdms04_scatplot}). Similarly, the ZIP detectors developed in the CDMS experiment benefit from amorphous silicon underlayers.

\subsection{Developments and perspectives}
Although not strictly necessary, a real 3-D determination of the interaction position in the detector would allow both a rejection of surface events and a much more precise control of the radioactive contaminants in the vicinity of the ({\it a priori} extremely pure) semiconductor\index{semiconductor} detectors. But the strongest motivation for a precise determination of the interaction position is the identification of heavy nuclear recoils from radon disintegration products at the surface of the detectors, or from materials facing them (see section "Event categories").

In future developments, the CDMS experiment intends to develop detectors with phonon sensors on both surfaces, in order to achieve this full 3-D position determination. For germanium detectors, where the speed of sound is lower than in silicon, massive detectors in the few hundred gram range may provide time arrival differences between the two sets of phonon sensors allowing a measurement of the interaction depth in the detector. Combined with the surface position determination already achieved in ZIPs, this would allow, at least at energies larger than a few tens of keV, a complete 3-D determination of the interaction profile. However, the time difference required to reconstruct the axial coordinate in the detector --a few microseconds-- appears difficult to be obtained at low energies, where most of the WIMP interactions are expected.

\begin{figure}
\includegraphics[width=.6\textwidth]{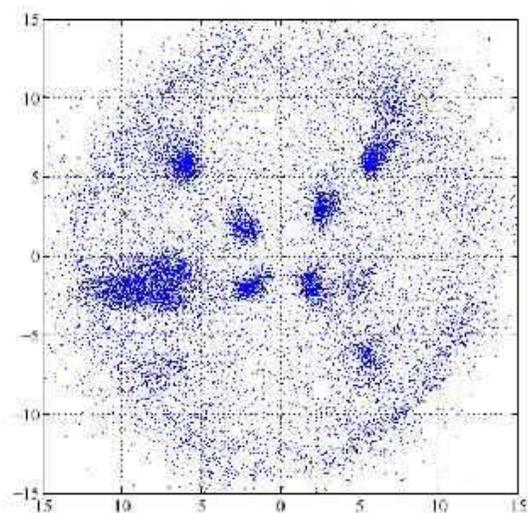}
\caption{ Two-dimensional position reconstruction capability of a ZIP detector of the CDMS experiment. The relative amplitudes detected by four phonons sensors are used to recover the x-y coordinates of an interaction in the detector. A series of radioactive point sources have been used to illuminate the detector and their contributions can be easily separated.}
\label{zip_2d}
\end{figure}

The determination of the interaction depth using the charge signal has been realized by the Orsay group~\cite{broniato01} for charge signal amplitudes $>\sim$ 50 keV electron equivalent (e.e.). But the signal/noise ratio is not sufficient to allow a precise position determination below $\sim$ 30 keV e.e. On the other hand, since a real 3-D position determination of the interaction is possible at higher energies, this feature could be used to discriminate signal from background events in future double-beta decay experiments~\cite{chardin03}. Additionally, the position reconstruction of interactions would lead to a better understanding of the radioactive contaminants present in key components of the next generation of WIMP detectors: sensors, cables, etc. Fig.~\ref{zip_2d} illustrates the present 2-D position reconstruction of the ZIP detectors developed in CDMS-II.

Due to the difficulty to develop Si detectors with purity levels as high as in germanium, the EDELWEISS experiment has chosen to use only germanium detectors. The contamination of nuclear recoil events by neutron interactions is then estimated by the coincidence rate between different detectors. Proportions of multiple interactions are still difficult to predict with precision, due to the systematic uncertainties in neutron simulations, but the quality of predictions gradually improves with high statistics neutron calibrations realized by the CDMS and EDELWEISS experiments. Future extended data takings in deep underground sites will allow to test and compare both neutron identification strategies.

\begin{figure}
\includegraphics[width=.6\textwidth]{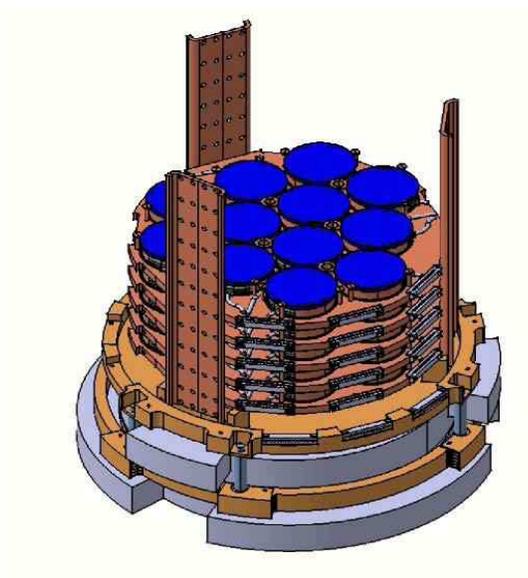}
\caption{ Drawing of the EDELWEISS-II detector set-up. Up to 120 Ge detectors of mass 320 g can be accommodated in a compact hexagonal arrangement at 10 mK. The lower plate, made of ultra-pure archeological lead, shields the detectors from the radioactivity of the dilution cryostat\index{dilution cryostat} (not shown)
\label{edelweiss2}}
\end{figure}

In conclusion, despite the problem of incomplete charge collection for surface events, charge-phonon detectors are still far from being limited by systematic uncertainties. They benefit from excellent crystal purities and energy resolution, allowing a precise comprehension of the background structure down to an energy of a few keV relevant for WIMP interactions. With present technology, it seems reasonable to assume that a 10$^{-8}$ picobarn sensitivity can be reached with cryogenic detector mass in the 50 kg range and a few years of exposure.

In the near future, both the CDMS-II and EDELWEISS-II experiments will greatly increase their low background exposures using detectors sets of total mass in the $\sim$ 10-30 kg range. These extended data takings will help define the improvements required to go beyond the sensitivity goal of this second generation of discriminating experiments.

\section{Event categories}
It is tempting, but incorrect, to consider that the problem of discriminating the alpha, gamma-ray and beta-induced radioactivity from WIMP- and neutron-induced nuclear recoil interactions is reduced to the problem of fitting two distributions, the distribution of nuclear recoils and the distribution of electron recoils. When the overlap between the distributions is limited, as in cryogenic detectors, or when no background subtraction is realized, the danger is limited, but when the discrimination capabilities are insufficient to clearly separate the nuclear and electron populations, at low energies or for intrinsic reasons, the sensitivity derived from a two-parameter adjustment of the spectrum observed in low-background data might be at considerable variance from the real populations. 

Even for single element targets, such as germanium or xenon, the experimentalist is confronted not only to the separation of electron recoils and germanium or xenon recoils but to several other populations. In particular, as noted previously, it is necessary to consider the heavy nuclear recoils from implantation\index{implantation} of radon daughter nuclei, or from the U/Th disintegration chains originating from either the crystals or the surrounding materials. For instance, the $\alpha$-disintegration of $^{210}$Po into the reaction:
$$\mathrm{^{210}Po} \rightarrow \mathrm{^{206}Pb} + \alpha$$
will obviously induce polonium recoils. Conservation of energy-momentum in the $\alpha$-disintegration shows that these heavy nuclear recoils have an energy equal to $\sim$ 103 keV for this particular reaction, and ranging from $\sim$ 70 keV to $\sim$ 110 keV for most other alpha disintegrations. These recoils can only be distinguished from WIMP interactions by detectors with excellent energy resolution and discrimination capabilities.

Yet another population must be taken into account with alpha tracks escaping the detector in the direction opposite to heavy nuclei. These alpha tracks have a quenching factor usually intermediate between electron recoils and heavy nuclear recoils for the same deposited energy. The energy released in the detector by the alpha particle depends on the implantation depth and the angle of emission with respect to the normal of the detector surface. It ranges from zero to the full alpha energy, leading to a continuous spectrum. Alpha tails, associated with alpha particles escaping from materials surrounding the detectors, such as copper supporting structures, provide yet another population since these stopping alpha particles have a much lower velocity and ionization power than the alpha particles leaving the detector after a few keV of energy loss. This is presently the most important background experienced by the CUORICINO\index{CUORICINO} experiment in its search for double-beta decay interactions.

Until now, we discussed the simple case of targets with a single nucleus (e.g. Ge, Xe, Ar). For targets containing several nuclei, the different responses of each participating nucleus must be taken into account. For example, in NaI targets, sodium and iodine recoils lead to rather different quenching factors, $\sim 1/3$ for Na recoils and $\sim 1/12$ for iodine recoils, complicating the interpretation of the observed energy spectrum. The situation is similar for bismuth germanate\index{BGO} (BGO) and calcium tungstate\index{CaWO$_{4}$} (CaWO$_{4}$) targets, where the detector response must be calibrated for each individual nucleus.

It results from this discussion that important systematic errors will result from the usual assumption of the existence of only two populations of electron and nuclear recoils. This is particularly true for experiments, using e.g. NaI and liquid xenon detectors, deriving their sensitivity limits after important background subtraction and with energy resolutions $>\sim$ 100 \% at the low energies relevant to WIMP interactions. The fact that the DAMA experiment has never reported the observation of the additional populations observed in the UKDMC~\cite{kudryavstev} and the Saclay NaI experiments~\cite{gerbier99} with a subset of the crystals used by the DAMA group, illustrates the systematic uncertainties of this experiment.

\section{Light-phonon detectors}
\subsection{Introduction}
Detectors able to detect simultaneously light and phonon\index{light-phonon} signals provide another powerful technique to distinguish a WIMP signal from the various components of the radioactive background. After initial developments at MeV energies~\cite{bobin,fiorini_scint}, two experiments, CRESST~\cite{bravin,meunier} and ROSEBUD~\cite{cebrian03}, are presently testing the phonon-scintillation discrimination scheme. Similarly to the charge-phonon detectors, the light output per unit energy is significantly different for nuclear and electron recoils. The ratio between the total energy, measured by the phonon signal, and the energy of the photons emitted in the interaction is then used to discriminate the nuclear and electron recoil interactions.

Various dielectric crystals have been tested by the CRESST and ROSEBUD experiments and have been shown to be luminescent\index{luminescence} at low temperatures with scintillation time constants less than a few millisecond. The fraction of energy effectively detected in the form of light is $\sim$ 1 \% for the best scintillators, which should be compared to the $\sim$ 10-20 \% fraction of energy converted into ionization in Ge and Si monocrystals. The existence of a variety of materials scintillating at very low temperatures has the important consequence that nuclear elements other than germanium and silicon can be considered as targets for WIMP interactions, which might prove essential to ascertain the existence of potential WIMP candidates.

\begin{figure}
\includegraphics[width=.6\textwidth]{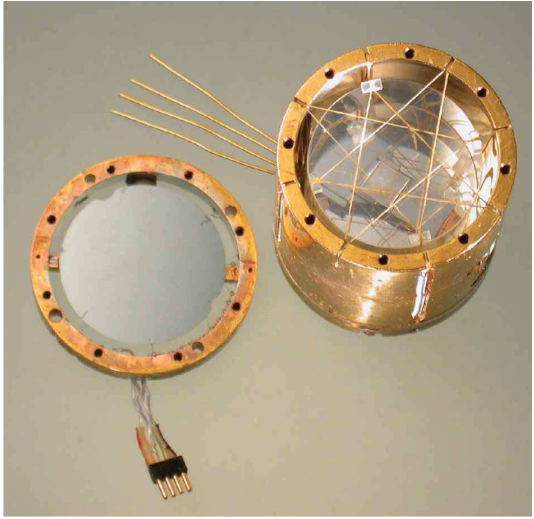}
\caption{ Photograph of a light-phonon detector of the ROSEBUD experiment. The phonon signal is determined by the temperature variation of a transparent 91 gram BGO crystal, while the light detector is constituted by a 25 mm diameter germanium wafer. Both detectors use NTD thermometers and are surrounded by a silvered copper reflector. This detector has been operated in the Canfranc underground laboratory (Spain)
\label{bgo_bolo}}
\end{figure}

Another important characteristic of the light-phonon detectors is the absence, at the present level of study, of any significant surface effects for electron recoil interactions, such as observed in charge-phonon surface interactions, where charge is badly collected. On the other hand, heavy nuclear recoils from radon disintegration products may represent an important background for these detectors since they are nearly indistinguishable from recoils of the heavier nuclei in the crystal itself (e.g. tungsten in CaWO$_{4}$ crystals, or bismuth in BGO crystals).

\subsection{Schematics of a light-phonon detector}
At the millikelvin temperatures required by the operation of massive phonon detectors, the use of conventional photodetectors is prohibited by their dissipation. The photon detector is therefore realized by a calorimeter with very small heat capacity, usually a very thin crystal wafer covered with a film absorber. The heat capacity of the light detector is minimized to allow the detection of the small number of photons emitted during the particle interaction and finally absorbed in the photon calorimeter. It should be noted that for a 1 keV electron recoil, only a few tens of photons are emitted, and that this number is further reduced by the quenching factor for a nuclear recoil and by the light collection efficiency.

This quenching factor was measured to be $Q \approx 1/7.4$ for oxygen recoils, and room-temperature measurements have established that tungsten nuclear recoils lead to quenching factors of $\approx 40$. Therefore, while oxygen recoils give rise to observable light signals down to $\approx 10$ keV, tungsten recoils usually lead to very small light signals, compatible with noise over most of the WIMP recoil energy interval.

Fig.~\ref{bgo_bolo} shows the principle of operation of a light-phonon detector developed in the ROSEBUD experiment~\cite{cebrian03}. This detector is constituted by a massive BGO (bismuth germanate) absorber facing a thin germanium wafer, 50-100 micron thick. To optimize light collection, both light and phonon detectors are placed inside a reflecting enclosure made of silvered copper. The temperature increase of the BGO crystal and the light detector are both measured by the impedance variation of a NTD (Neutron Transmuted Doped) sensor.

The detectors developed by the CRESST experiment are based on a similar principle, using a CaWO$_{4}$ massive detector facing the light detector constituted by a $30 \times 30 \times 0.4$ mm$^{2}$ silicon wafer. The main difference lies in the sensor technology: instead of NTD sensors, TES (Transition Edge Sensors) made of tungsten are used to measure the temperature increase of both the phonon and light channels of the detector. The detectors are operated at a temperature of $\approx$ 10 mK, in the middle of the superconducting to normal conducting state of the tungsten sensor. Sensors constituted by Ir/Au bilayers have also been realized and present similar but somewhat lesser sensitivities. TES sensors benefit presently from a better sensitivity than NTD sensors.

\subsection{Discussion}
For dielectric materials with reasonable Debye temperatures, the measurement of the phonon signal, i.e. of the major part of the total energy deposited in a particle interaction, has already been realized on massive detectors in the several hundred gram range such as CaWO$_{4}$, BGO or TeO$_{2}$. Energy thresholds of a few keV are now routinely achieved on such massive detectors by several groups.

Detection of the light signal appears more challenging as it requires several constraints on the target crystal. The scintillation efficiency must be sufficiently large for the light signal to emerge from the thermal noise. Moreover, the high energy cost of a single photon has for consequence that only a small fraction of the energy of a particle interaction is converted into photons for low-energy nuclear recoils. The crystal must then have a good transparency at the fluoresced wavelength to ensure a homogeneous signal response within the detector. Also, crystals must be selected that present relatively fast scintillation time constants at very low temperatures, although this constraint is partially attenuated by the relative sluggishness of the thermal detectors, with thermal relaxation times usually in the millisecond range. In addition, the quenching factor --here the scintillation efficiency for a nuclear recoil compared to electron recoils of the same energy-- must be sufficiently different from unity to provide a clear identification of nuclear recoils down to low energies, but sufficient for light and phonon signals to be detectable for nuclear recoil events.

Only a few tens of photons are emitted in a 1 keV electron recoil in a CaWO$_{4}$ or a BGO crystal. This number is further reduced by the quenching factor when the interaction is due to a nuclear recoil. For CaWO$_{4}$, the quenching factor is $\sim1/7.4$ for oxygen recoils, and the number of photons per keV of recoil energy is limited to a few units per keV. The scintillation quenching factor is presently not measured for germanium and bismuth recoils in BGO, but phenomenological models lead to expect quenching factors for these nuclei significantly smaller than for oxygen recoils. A neutron beam test facility to measure quenching factors using tagged mono-energetic neutron beams has been developed by the CRESST and EDELWEISS experiments~\cite{tecnomusiq}. This facility is a fundamental tool to calibrate detector responses to electron and nuclear recoils.

\subsection{First results}
Initial data takings by the CRESST-II experiment, using two CaWO$_{4}$ 300 g detectors, exhibited a population of phonon-only events, for which the amplitude of the light signal was compatible with zero. The event rate, a few events per kg and per day, was a priori too high to be attributed to WIMP interactions, as it would contradict both the CDMS and EDELWEISS results, or to neutron interactions, predicted to be a factor $\approx$ 10 smaller. These events were therefore attributed to dissipative events related to stress on the detector crystals and were suppressed by replacing the teflon supports holding the crystal with copper-beryllium springs. Another improvement on the detectors was achieved by polishing the surface of the detectors to reduce the effect of total internal reflection and provide a better homogeneity of the light response and resolution. The phonon channel resolution at the 46.5 keV $^{210}$Pb $\gamma$-ray peak is improved to $\approx$ 1 keV FWHM. Finally, a polymeric foil with very high reflectivity~\cite{weber} ($\approx 99 \%$ at 420 nm) was used to optimize the scintillation light collection.

Further data takings using the modified detectors showed a residual population of events compatible with oxygen nuclear recoils, with a small but non-zero light signal. The absence, for the CaWO$_{4}$ detector with the best resolution, of any phonon-only event between 20 keV and 40 keV shows that light-phonon detectors can be used to discriminate oxygen recoils from calcium and tungsten nuclear recoils, the latter recoil population providing negligible light emission. Presently limited by the neutron background, expected to induce a nuclear recoil event rate compatible with the observed rate, CRESST presents in its latest result~\cite{angloher04} a sensitivity comparable to that of EDELWEISS. In its upgraded phase with a polyethylen neutron shield and a plastic scintillator muon veto, starting at the beginning of 2005, the CRESST-II experiment, using a total of 10 kg of CaWO$_{4}$ crystals, is expected to improve its sensitivity down to the level of a few 10$^{-8}$ picobarn.

\begin{figure}
\includegraphics[width=.6\textwidth]{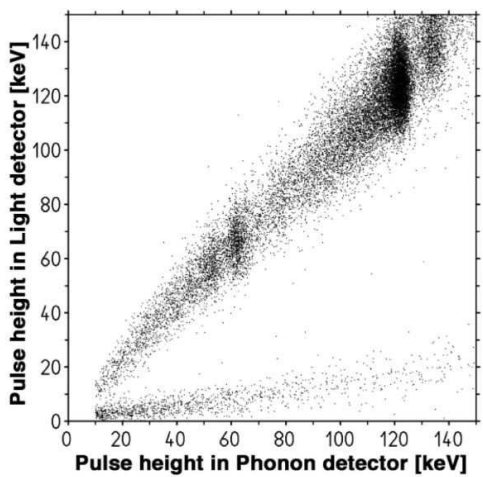}
\caption{
\label{cresst_scatplot}
Projection in the ($E_{Recoil}$, Q) plane of the events recorded in a 15 g CaWO$_4$ light-phonon prototype detector of the CRESST experiment during a $^{252}$Cf neutron calibration. A clear separation of the electron recoil and nuclear recoil populations can be observed down to an energy of $\sim$ 15 keV recoil energy. No intermediate population, due e.g. to surface interactions, is apparent in this light-phonon discrimination scheme.
}
\end{figure}

Fig.~\ref{cresst_scatplot} shows the quenching factor distribution obtained by the simultaneous measurement of light and phonon signals for a CaWO$_4$ crystal irradiated by $\gamma$-ray and neutron sources. It can be seen that the separation of the two populations of electron recoils and oxygen recoils is excellent down to $\sim$ 15 keV recoil energy, sufficient in principle to detect a significant fraction of WIMP interactions with mass compatible with accelerator constraints. However, it should be noted that only oxygen recoils are observed in Fig. \ref{cresst_scatplot}, while most of the WIMP-nucleon interaction cross-section will be, at least for a scalar coupling, associated with the heavier tungsten and calcium nuclei (respectively germanium and bismuth for BGO). With the sensitivities currently reached by the present CRESST detectors, interactions on these heavier nuclei would lead to a light signal hardly emerging from thermal noise.

For this reason, the number of phonon-only events, associated with spurious heating of the detector due, for example, to micro-fractures in the crystals, dissipation associated with vibrations in its supporting structure, or electromagnetic spurious heating, must be reduced to a negligible level in order to compete with the cryogenic charge-phonon experiments. Phonon-only events have been observed by the CDMS-I and EDELWEISS-I experiments, using germanium detectors, and in a first version of the CRESST experiment, using sapphire detectors. These phonon-only events are not expected to represent a problem for charge-phonon detectors down to sensitivities $< 10^{-8}$ picobarn since it is possible for these experiments to reject events lacking a simultaneous ionization signal. In addition, for the new generation of ZIP~\cite{clarke} detectors used in the CDMS-II experiment, these phonon-only events will {\it a priori} not be detected since only phonons with an individual energy sufficient to break Cooper pairs\index{Cooper pair} of the Al fins, well above the detector temperature of $\sim$ 20 mK, are detected.

Clearly, the light-phonon technique is still in development and further improvements on light collection and sensitivity to WIMP interactions are expected over the next few years.

\subsection{Conclusions}
The present performances of light-phonon detectors offer an excellent discrimination between electron and nuclear recoils, a key feature for WIMP detection. On the other hand, for all scintillators at low temperatures presently tested, the small quenching factor for light emission by heavy nuclear recoils (such as bismuth in BGO, or tungsten in CaWO$_{4}$) leads to a challenging light detection for these nuclei, which represent a very large fraction of the cross-section for spin-independent WIMP interactions. At present, the fact that the signal amplitude of the light channel is still compatible for these nuclei with thermal noise imposes to accept dissipative (``phonon only'') events as WIMP interaction candidates. The number of these events must obviously be kept negligible. Also, heavy nuclear recoils from radon disintegration products will be indistinguishable from WIMP-induced heavy nuclear recoils in the target crystal. To fight this dangerous background, CRESST has used scintillating coatings of the light reflectors to detect the associated alpha particle escaping the detector. Using this veto scheme, CRESST-II is able to reject a large fraction ($\sim 80\%$) of these radon-induced nuclear recoils.

Despite these constraints, the present sensitivity of the CRESST-II experiment is already at the level of the CDMS and EDELWEISS sensitivities, while still mostly limited by its neutron background. Therefore, it appears extremely important to pursue this line of development since further improvements in the light collection efficiency might lead to the possibility to detect the light signal of higher mass nuclei. Also, some presently untested scintillators may be found to present a more favorable light yield for nuclear recoils. Solving the present difficulties of this detection scheme will allow using additional target materials, an essential feature to ascertain the WIMP origin of a signal.

\section{Other cryogenic detectors}

\subsection{MACHe3 : $^{3}$He, the purest of all materials}
Germanium and silicon monocrystals benefit from extremely low levels of impurities. In detector quality crystals, the difference in acceptor and donor impurities can be as low as a few $10^{9}$ per $\mathrm{cm^3}$, and the concentration of radioactive impurities is much lower than this already impressively small number, assuming the cosmogenic production of radioactive isotopes has been limited. But superfluid\index{superfluid} helium-3\index{helium-3} and helium-4\index{helium-4} surpass these semiconductor materials, with an impurity concentration practically reduced to zero. At subkelvin temperatures for superfluid $^4$He, and below 1 millikelvin for $^3$He, essentially no impurities can remain in a stable way within these liquids: they just sink at the bottom of the helium container (or float for electrons and hydrogen). At a temperature of 100 $\umu$K, not even a single atom of the chemical twin $^4$He is able to remain diluted at equilibrium in a kilogram of superfluid $^3$He!

This extreme purity has been a motivation for proposing a dark matter detector where the temporary intrusion of an external particle creates quasiparticles which are detected by the attenuation of a vibrating wire inside the liquid. Energy thresholds of a few keV have already been achieved with small $^3$He samples, and a multicell detector has been proposed~\cite{bradley,santos}. Two drawbacks, however, are reducing the attractiveness of this elegant detector: the cost of $^3$He is of the order of 1000 euros per gram, and there is no discrimination proposed against the very low energy X-ray background which may mimic WIMP interactions. Present event rates per unit mass and time are several orders of magnitude higher than the rates reached by the CDMS, CRESST and EDELWEISS experiments.

\subsection{ORPHEUS: superconducting superheated granules}
The ORPHEUS experiment, located in a shallow underground laboratory in Bern (Switzerland), has developed a digital cryogenic detector, using superconducting tin granules\index{granule (superconducting)} in a metastable state~\cite{bernas,orpheus}. At very low temperatures, deposition of a few keV of energy is able to trigger a transition from the superconducting to the normal state for individual tin granules with diameters of 28 $\umu$m and 36 $\umu$m. The detection of the transition of a single grain is made possible by measuring the magnetic flux change induced by the disappearance of the Meissner effect\index{Meissner effect}, expelling magnetic field lines from a superconductor. The digital character of this detector is then related to the fact that the magnetic field change in a single granule transition is only related to the transition itself and largely independent of the deposited energy.

A total mass of 450 g of tin granules set in 56 $\times$ 8g modules at a temperature of $\sim$ 115 mK has been used in the ORPHEUS experiment. It should be noted that diamagnetic\index{diamagnetic} interactions\index{diamagnetic interaction} between the individual granules impose a reduced filling factor for this detector: tin granules are mixed with teflon powder to reduce these interactions and $\sim$ 10\% of the total mass of a module is active in the present stage of the experiment. This parameter is particularly important as it directly relates to the possibility of rejecting an important fraction of the Compton gamma-ray background, expected to produce a large fraction of multiple site energy depositions.

Preliminary results have been presented by ORPHEUS~\cite{orpheus_idm02}, showing background event rates in the 2 $\times 10^3$ events/kg/day range, with only a limited fraction of background events rejected by the multiplicity of grain transitions. Although in principle the background energy spectrum of this detector can be determined by varying the temperature and magnetic field, inhomogeneities in individual grain response and diamagnetic interactions make a precise determination of the background spectrum unlikely. The present sensitivity of the ORPHEUS experiment to scalar WIMP interactions is $\sim 5 \times 10^{-1}$ picobarn, and therefore $\sim$ 4 orders of magnitude above that of the discriminating cryogenic experiments.

\subsection{The Tokyo LiF bolometer experiment}
The Tokyo group~\cite{ootani,miuchi} has used lithium fluoride bolometers at a temperature of $\sim$ 10 mK to take advantage of the excellent axial WIMP-nucleon cross-sections of the fluoride nucleus~\cite{ellis_flores} and, to a lesser extent, of the lithium nucleus. After an initial experiment in a shallow laboratory, close to surface, which used a set of 8 $\times$ 21 g LiF detectors~\cite{ootani}, a second experiment was performed in the Kamiokande underground laboratory to benefit from the reduced cosmic-ray induced background~\cite{miuchi}. Surprisingly enough, the Kamiokande measurements have not led until now to improvements in background performances and the sensitivity reached by the Tokyo LiF detectors still falls short by nearly three orders of magnitude of the sensitivity required to sample the most optimistic SUSY models. While this experiment provides an illustration of the fact that a large variety of target materials can be used in bolometers, it also confirms that most of these detectors, lacking background discrimination capabilities, are unable to provide sensitivities comparable to indirect detection experiments such as Superkamiokande~\cite{superk} and Amanda~\cite{amanda}. These experiments, searching for high-energy neutrinos from the core of the Sun, provide much more stringent constraints for spin-dependent WIMP-nucleon couplings than those obtained by these non-discriminating experiments~\cite{kamionkowski}.

Recently, the ROSEBUD experiment has shown that LiF crystals emit light under particle interactions, and offer the possibility to distinguish electron recoils from alpha particle and nuclear recoil interactions by their different light/phonon ratio~\cite{marcillac03}. However, the light yield in these interactions appears until now too limited to be used to efficiently discriminate low energy nuclear recoils from the gamma-ray background. Still, this development could lead to use LiF bolometers as sensitive neutron detectors.

\begin{figure}
\includegraphics[width=.6\textwidth]{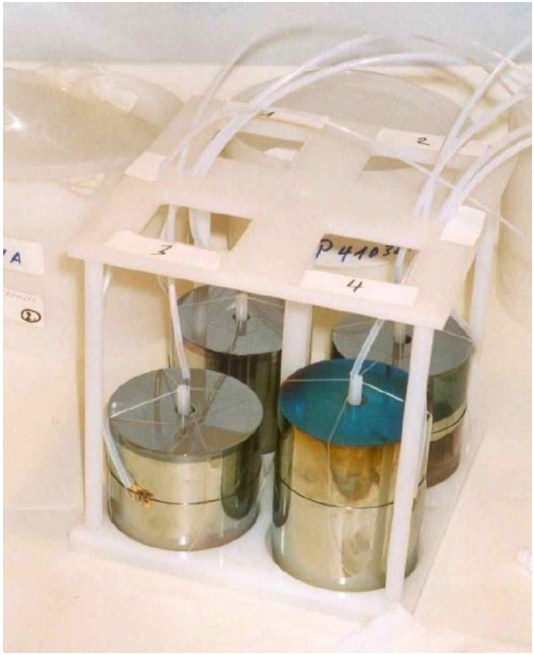}
\caption{ Photograph of a detector subset of the GENIUS-TF (Test Facility) experiment, comprising 14 Ge detectors, with a total active mass of $\sim$40 kg. These naked detectors, with minimal material other than germanium, are operated in liquid nitrogen, which can be continuously recirculated and purified. This experiment, dedicated in its first stage to Dark Matter search, is presently in operation in the Gran Sasso underground laboratory
\label{genius_tf}}
\end{figure}

\subsection{Detecting the recoil direction}

Assuming a non-rotating halo relative to the galaxy, WIMPs will exhibit a strong anisotropy with respect to the laboratory frame, rotating with a daily period and reflected in nuclear recoil directions induced within a detector~\cite{directional}. Therefore, several groups~\cite{rich88,buckland,lanou,bandler,collar,lehner,snowden_ifft95,snowden_ifft97,drift00} have addressed the challenging experimental question of determining the recoil direction under a WIMP interaction in order to use this strong directional signature. Compared to the annual modulation signature, which presents at best a modulated amplitude of 7\%, the recoil direction, at least for nuclei with a mass comparable to that of the incident WIMP, might present much larger anisotropies, of the order of 50 \%. In addition, the directional signature will increase with the recoil energy and is therefore much less sensitive to threshold effects than the annual modulation technique. The difficulty lies of course in the possibility to reconstruct the recoil direction in an interaction of a few tens of keV at most.

\begin{figure}
\includegraphics[width=.6\textwidth]{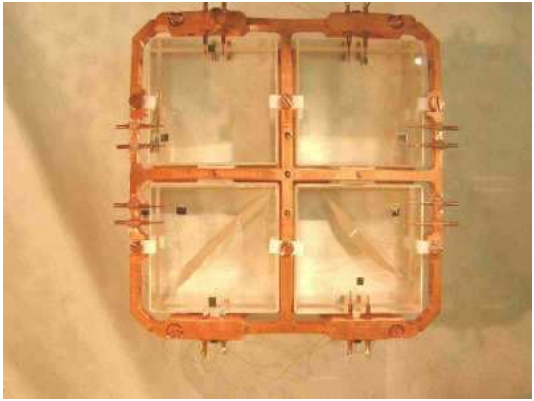}
\caption{ Top view of the detector tower of the CUORICINO experiment. Each plane of the tower comprises four detectors, weighing 760 g each, except for two of the lower planes, each comprising 9 $\times$ 340g. The energy of an interaction is measured by the impedance variation of a Neutron Transmuted Doped (NTD) Ge sensor. These detectors are operated at a temperature of $\sim$ 10 mK in the Gran Sasso underground laboratory (Italy)
\label{cuoricino}}
\end{figure}

For such small deposited energies, the nucleus typically recoils by sub-micron sized distances in usual solid state detectors. To keep track of the anisotropy information of these low energy interactions, an original method has been proposed in the context of the HERON\index{HERON} project~\cite{lanou} for real-time solar neutrino detection. In the remarkably homogeneous medium of superfluid helium, a local deposition of energy with a high density will create an opaque roton cylinder around the track. Due to Lambert's law, the resulting roton emission is then anisotropic and can be correlated to the incident particle direction~\cite{bandler}. However, this beautiful experiment has only been demonstrated with alpha, gamma and electron sources with energies larger than 300 keV, and appears to be inoperative at the much smaller recoil energies characteristic of WIMP interactions.

Therefore, at present, the most promising technique to identify the recoil direction under a WIMP interaction is under development in the context of the DRIFT (Directional Recoil Identification From Tracks) project~\cite{snowden_ifft97,drift00}, at room temperature, using low density gas targets such as CS$_2$.

\section{Towards ton-scale Dark Matter experiments}

The best sensitivity to WIMP interactions is presently obtained by the CDMS experiment\cite{cdms04}, using four Ge detectors of total mass 1 kg and fiducial mass $\sim$ 0.5 kg. It corresponds to a maximum sensitivity of $\sim$ 4 10$^{-7}$ picobarn, assuming standard halo parameters. The CRESST\cite{cresst04} and EDELWEISS\cite{edelw02,sanglard04} experiments provide similar, although somewhat lesser sensitivities. Although some of the most optimistic SUSY models are tested by these experiments, models favored by SUSY theorists correspond to cross-sections $<\sim$ 6 $\times 10^{-8}$ picobarn, requiring one order of magnitude increase in sensitivity to be tested. A first part of these models is expected to be sampled by the next generation of cryogenic experiments, such as CDMS-II, CRESST-II and EDELWEISS-II, using target masses in the 10 to 35 kg range.

Reaching a sensitivity to WIMP cross-sections in the 10$^{-10}$ picobarn range, allowing to test a much larger fraction of SUSY models, requires ton-scale targets: a cross-section of 10$^{-10}$ picobarn would provide $\sim$ 10 events per ton and per year for germanium and xenon targets, assuming full detection efficiency above 15 keV recoil energy. Naively, a larger number of events would be expected for heavier nuclear targets such as xenon, iodine or tungsten than for germanium, but the loss of coherence at energy transfers larger than a few tens of keV saturates the number of events as a function of the atomic number {\it A} of the target. Therefore, the above nuclear targets all provide similar WIMP event rates per unit of target mass. Reaching the 10$^{-10}$ picobarn grail will obviously be extremely difficult even for ton-scale experiments.

\subsection{Ton-scale experiments without background discrimination}
Three groups have proposed ton-scale Dark Matter experiments involving detectors with limited or no background discrimination. The DAMA/LIBRA and GENIUS-TF/GENIUS experiments use detectors at room temperature and liquid nitrogen temperatures respectively, while the CUORICINO/CUORE experiment is using cryogenic detectors at a temperature of $\sim$ 10 mK.

\subsubsection{The DAMA/LIBRA experiment}

The DAMA group, presently operating the LIBRA NaI detector with a total detector mass of 250 kg, has proposed to increase its sodium iodide detector mass to one ton~\cite{dama_oneton}, with the objective to confirm its evidence for an annual modulation. However, the present systematic limitations of this experiment, which already accumulated an exposure of $\sim$ 100 ton $\times$ day, make it improbable that significant improvements in sensitivity will be reached using this technique.

\subsubsection{The GENIUS-TF/GENIUS experiment}

The GENIUS project~\cite{genius}, mainly devoted to the study of double-beta decay, is proposing two stages with 1 ton and 10 tons of enriched $^{76}$Ge detectors. This project has based its approach on the excellent purity of germanium crystals. The analysis of data accumulated over several years by the Heidelberg-Moscow experiment indicated that the most prominent contributions to the background originated from the copper structure surrounding the detectors. Therefore, GENIUS intends to reduce all detector materials other than germanium to a minimum, with a proposed shielding consisting of a 13 m diameter tank of liquid nitrogen, constantly recirculated. Prototype detectors have been operated successfully in liquid nitrogen with FETs deported at distances of up to several meters, with energy resolutions comparable or even exceeding those obtained with conventional detectors. According to the GENIUS proponents, the main limiting factors are therefore the radioactivity of all other materials than the germanium target. The internal radioactivity of germanium itself, resulting from cosmogenic activation, and most notably $^{68}$Ge, is then the second most prominent background. The success of GENIUS requires that all materials other than liquid nitrogen and germanium (wiring, charge collection electrodes, etc.) are of limited mass and maintained radioactively ultrapure.

However, to reach the GENIUS target sensitivity of 2 $\times$ 10$^{-9}$ picobarn, the radioactive background at low energies must be reduced by more than three orders of magnitude compared to the best present background levels, reached by the IGEX and the Heidelberg-Moscow experiments after several years of development. This level of extrapolation requires an intermediate test, which is presently attempted with the GENIUS-TF facility~\cite{genius_tf}, using a total of 14 HPGe-detectors with a total active mass of $\sim$ 40 kg. The cosmogenic activation of $^{68}$Ge and $^{71}$Ge, and the ubiquitous presence of tritium also represent two particularly dangerous low energy backgrounds for the GENIUS project. Underground crystal fabrication is explored, as this would reduce the $^{68}$Ge activation by one to two orders of magnitude (depending on the fabrication site and transport duration to the deep-underground experimental site).
Also mainly aimed at the detection of double-beta decay, the MAJORANA project~\cite{majorana} will involve 500 kg of enriched $^{76}$Ge and intends to test the existence of WIMPs down to cross-sections $\sim 10^{-8}$ picobarn through the annual modulation signature.

\subsubsection{The CUORICINO/CUORE experiment}

The CUORICINO experiment~\cite{cuoricino}, test stage of the CUORE\index{CUORE} experiment~\cite{cuore}, is using a compact structure of $44 \times 760$~g and $18 \times 340$~g TeO$_{2}$ bolometric detectors, operated at T~$\sim$~10 mK, with a total target mass of $\sim$~40~kg (Fig.~\ref{cuoricino}). The main scientific aim of the CUORICINO/CUORE experiment is, similarly to GENIUS, the observation of 0-$\nu$ double-beta decay. This experiment also aims at the observation of a WIMP signal through the challenging annual modulation signature~\cite{cuore,ramachers_annual,hasenbalg}. The cosmogenic activation is not a problem here since radioactive isotopes of oxygen and tellurium are short-lived. The main problem encountered by the CUORICINO experiment appears to be related to U/Th contamination in the copper supporting structure, and to the surface implantation of heavy nuclear elements in crystals resulting from radon disintegration products. While an important part of this background is observed at MeV energies, these surface implantations lead in all experiments to a continuous background extending down to energies characteristic of WIMP interactions, through the escape of alpha tracks and by the low energy surface beta contaminations (from, e.g., $^{210}$Bi and $^{210}$Pb $\beta$-decays).

Therefore, the techniques presently developed in the main non discriminating WIMP direct detection experiments require an extrapolation by more than three orders of magnitude in their background performances to reach their target sensitivities of a few 10$^{-9}$ picobarn.

\subsection{Large-scale cryogenic discriminating experiments}

A different strategy is proposed by the cryogenic
experiments. The emphasis here is not so much on the radioactive
purity, obviously still necessary, but on the quality of the discrimination between electron and nuclear recoils to identify a possible WIMP signal. During
the next few years, the CDMS-II, CRESST-II and EDELWEISS-II
second-generation experiments will each use a mass of detectors in the 10 kg range to test
the various discrimination
strategies (simultaneous measurement of fast phonons and charge,
thermal phonons and light, or thermal phonons and charge) while
already testing a significant part of the SUSY allowed models.

To reach the goal of a one-ton cryogenic experiment, cryogenics does not appear {\it per se} as the main challenge. In this respect, cooling down a 2.3 ton Al antenna at a
temperature of 100 mK has already been achieved by the NAUTILUS
cryogenic search for gravitational waves~\cite{nautilus}, while in future years, the GRAIL project intends to develop a 150-ton gravitational antenna at a temperature of $\sim$ 10 mK~\cite{grail}.

On the other hand, cryogenic experiments still have to demonstrate that they
can reliably operate very large numbers of detectors. As mentioned previously, the CUORICINO experiment~\cite{cuoricino} is presently successfully operating a total of $\sim$ 40 kg of crystals, representing the present largest
cryogenic experiment for dark matter and double beta decay
search. In a few years, the recently approved CUORE project~\cite{cuore} intends to use a
thousand 760 g $\mathrm{TeO_2}$ crystal detectors at a temperature of 10 mK,
almost reaching the one ton objective. Although the CUORE
project, principally aimed at the neutrino mass measurement, will
hardly be competitive with future dark matter searches due to its
present lack of discrimination capabilities, this experiment
shows that large numbers of
detectors at very low temperatures can be considered and
realized.

For such large-scale experiments, ease of fabrication
and reliability represent two essential factors. In this respect,
thin film and magnetic sensors might offer an elegant solution to the problems of
reproducibility and channel multiplicity. Several groups~\cite{bravin,irwin,cryo_review2,marnieros} have developed such MIT, TES and magnetic sensors with sensitivities comparable
or even exceeding that of conventional thermal sensors such as Neutron
Transmutation Doped sensors (NTDs), used in the CUORE and EDELWEISS experiments.

\subsection{Strategy for ton-scale dark matter cryogenic experiments}

Aside from the difficulty of integrating a thousand cryogenic detectors at temperatures in the 10 millikelvin range, ton-scale discriminating cryogenic experiments must address the identification of neutron background, which will mimic WIMP-induced nuclear recoils. In addition, an increase of the individual detector mass appears desirable, while maintaining an excellent energy resolution required by a particle identification on an event-by-event basis.

\subsubsection{The neutron background}
Three main neutron components are present in a deep underground laboratory, required to operate a Dark Matter direct detection experiment below the present sensitivity to WIMP interactions of $\sim 10^{-6}$ picobarn reached by the CDMS, CRESST and EDELWEISS experiments.

\begin{figure}
\includegraphics[width=.6\textwidth]{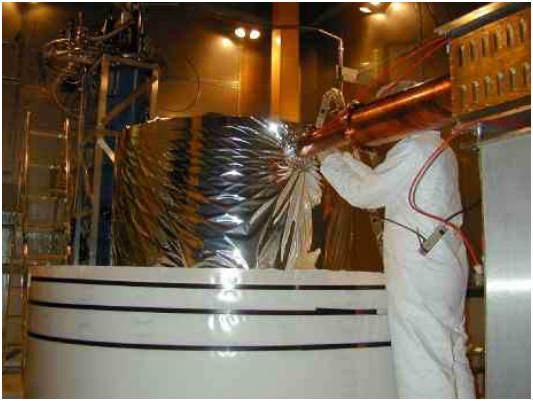}
	\caption{ Setup of the CDMS-II experiment in the Soudan underground laboratory (Minnesota). The cold box, made of ultrapure copper, is surrounded by a passive polyethylen shielding against neutrons from the rock and concrete. An external lead shield is used as a further protection against the gamma-ray background. In the Soudan laboratory, the muon flux is reduced by a factor $\sim 10^{5}$ compared to ground level, but a muon veto (not shown on the photograph) is used to identify the remaining muons or showers crossing the detector setup}
	\label{cdms_setup}
\end{figure}

The first component comes from the spontaneous fission\index{fission} of uranium and from spallation\index{spallation} reactions in the rock and concrete walls of the underground laboratory shielding the experiment from the muon-induced background. For both fission and $\alpha$-n reactions, neutrons are emitted at typical energies of a few MeV. For the best sites, neutron fluxes are of the order of $10^{-6} \mathrm{neutrons/cm^{2}/s}$. This neutron flux component can be efficiently reduced by several orders of magnitude, at levels corresponding to WIMP cross-sections less than 10$^{-9}$ picobarn, by using typically 60 cm of low-Z shielding, such as polyethylene. The protection setup used by the CDMS-II experiment in the Soudan underground laboratory is shown in Fig.~\ref{cdms_setup}. Without this protection, this neutron
flux component has already been detected by discrimination
experiments like EDELWEISS and CRESST, operating in deep-underground sites. It is presently the main background limiting the sensitivity of the first stage of the CRESST-II experiment, without neutron protection.

A second neutron component is associated with the neutron
production by muons crossing the copper and lead shieldings, used as gamma-ray protections but acting here as 
neutron multipliers. This neutron background can be effectively
reduced to a negligible level by identifying and vetoing the
muons crossing the protective setup around the detectors. The CDMS-II, CRESST-II and EDELWEISS-II cryogenic experiments are using plastic scintillator veto to identify muons and reject this fast neutron background. Different strategies are followed by EDELWEISS and CDMS in their control of muon-induced neutron background : in EDELWEISS, the polyethylene shielding is external to the low-radioactivity lead shielding, favoring the neutron production associated with muons close to the detectors, while the reverse solution has been chosen by CDMS. The higher muon flux in the Soudan underground laboratory, by a factor $\sim$ 40 compared to the Frejus underground laboratory, strongly constrains the solution chosen by the CDMS experiment, unless an extremely efficient muon veto is used.

Finally, even under the protection of thousands meters of water
equivalent, high-energy muons with energies in the TeV range may
lose in a catastrophic way a large fraction of their energy in a
single deep-inelastic scattering interaction on a nucleus. Typically 10
\% of the long range hadronic component is then carried away by several thousand 
neutrons, at very low shower densities, making this fast neutron
background extremely difficult to detect. A large active scintillator
shield, as used, for example, in the
Borexino or the KARMEN neutrino experiment~\cite{borexino,karmen} can strongly reduce this small but dangerous background.

Beyond these three neutron backgrounds, it is probably illusory to believe that it will be possible to reduce at a level below a few neutrons per year the internal neutron emission, by U/Th and alpha-n reactions, by a Dark Matter setup of total mass $>$ 100 tons, and particularly of the $\sim$ 10 tons of various materials (cryostat, shielding, cabling, electronics) in the immediate vicinity of the detectors.

The residual neutron background can be efficiently monitored by
studying the multiple scatter interactions in an array of densely
packed detectors, since fission and spallation neutrons will suffer tens of scatterings before reaching energies of a $\sim$ 0.1 MeV, where their scatterings become undetectable. For example, the compact detector arrangement presently studied in the CUORICINO experiment (Fig.~\ref{cuoricino}) involves a series of 760 g detectors in direct close view of each other. Similarly, EDELWEISS-II will be using up to ten planes with hexagonal paving of 12 Ge detectors. In the CUORE detector setup, the rejection
factor by identification of multiple scatter neutron events for the 512 inner detectors will reach two orders of magnitude. Such a compact arrangement of detectors will be necessary to clearly identify the residual internal neutron background. Using such a multiple scattering strategy, the neutron background can probably be kept below the level of a few $10^{-5}$ evt/kg/keV/day, corresponding to WIMP interaction cross-sections of $\sim 10^{-10}$ picobarn.

A different strategy has
been used by the CDMS experiment, with interspersed germanium
and silicon detectors to measure the neutron flux on two different
target materials. For WIMPs with mass compatible with accelerator constraints and predominant scalar couplings, Ge is $\sim$ 10 times more sensitive per unit of mass than Si to WIMP interactions, and it is therefore expected that Si detectors will mainly detect the remaining fast neutron flux, at the expense of a reduction of the fiducial mass for WIMP detection.

\subsubsection{Detector mass and resolution}
Similarly to CUORE, which is using 760 g detectors, it appears important to increase the individual detector mass of discriminating experiments, presently in the 250-320 g range, to $\sim$ 1 kg. Going much beyond this mass will degrade the energy resolution, while the cost of individual crystals will increase rapidly. For example, it seems difficult at present to obtain high purity Ge crystals with diameter significantly larger than 100 mm. For a 1 kg detector, this corresponds to a thickness of $\sim$ 30 mm. Further increasing the thickness is possible, but will reduce the electric field used to collect charges in the crystal since the collecting voltage must not exceed a few volts due to the Neganov-Luke effect.
With such 1-kg detectors, the energy resolution at WIMP interaction energies (a few keV) can be kept below 1 keV : EDELWEISS 320 g and CDMS 250 g Ge detectors have shown energy resolutions $\sim$ 300 eV. With the optimization of sensor parameters, energy resolutions better than 500 eV are expected for these 1 kg detectors. These performances should be compared to the energy resolution of present liquid xenon detectors, where a 100\% energy resolution is only obtained at $>$ 50 keV recoil energy ($\sim$ 8 keV e.e.)
Similarly, the CRESST and ROSEBUD experiments are developing detectors with individual mass in the 1-kg range with resolutions in the keV range, and the comprehension of the background structure at low energies appears as a key feature in favor of cryogenic experiments.

\section{Conclusions}
Three cryogenic experiments, CDMS, CRESST and EDELWEISS, presently provide the best sensitivity to WIMP interactions, one order of magnitude better than their present competitors and still far from being limited by systematic uncertainties. Despite the complexity of their operation, they provide a much better background control than other competing experiments, presently limited by much poorer energy resolutions and background discrimination performances. Obtained with small prototype detectors, in the few hundred gram range, these performances are susceptible to a further sensitivity increase of nearly two orders of magnitude during the next few years. Starting from the present sensitivity to cross-sections of $\sim 10^{-6}$ picobarn, which probes a first region of optimistic SUSY models, the present techniques of charge-phonon and light-phonon detectors should allow to test SUSY models down to cross-sections of $\sim 10^{-8}$ picobarn, a region of SUSY parameter space considered as more realistic by particle theorists~\cite{ellis00}.

Cryogenic experiments at the 10-kg scale will start being operated early in 2005, but exploration of most of SUSY models will require reaching the fantastic sensitivity of $\sim 10^{-10}$ picobarn. At this level, the integrated number of events in the relevant energy range is only a handful per ton of detector and per year. Therefore, based on existing technology, experiments in the 100 kg to 1 ton range are already designed in the EURECA (European Underground Rare Event search with Calorimeter Array) and SDMS projects, although new developments in background rejection will probably be required to reach the limit sensitivity of ton-scale experiments.

Particularly important, among the developments that might mark a further progress in discrimination performances, are the identification with high efficiency of surface events, and most notably heavy nuclear recoils associated with surface radon decay products. Light-phonon detectors, provided they are able to keep to a negligible level their dissipative ("phonon-only") events, which might otherwise be confused with heavy nuclear recoils, will provide an alternative set of nuclear targets, essential to finally identify a potential WIMP signal. For these three techniques, in order to reach the one-ton scale at a reasonable cost, it will be important to increase the individual detector size to $\sim$ 1 kg. 
A full 3-D position determination of the interaction down to low energies would also allow to check the homogeneity of the WIMP candidates events together with the location of the remaining radioactivity, but this capability appears as a daunting technical development below $\sim$ 10-20 keV recoil energy.

Finally, developments realized for dark matter direct detection with cryogenic detectors may also be beneficial to other physics applications. In particular, charge-phonon detectors might allow to identify and reject the two main radioactive backgrounds of present double beta decay experiments: surface alpha interactions and Compton $\gamma$-ray interactions. Other applications are the observation of coherent neutrino scattering, which can be easily achieved in future neutrino factory beams with present charge-phonon and light-phonon detectors, a high sensitivity search for the neutrino magnetic moment, and possibly real-time solar neutrino detection using the ZIP position resolution in the several hundred keV range.

In the same way that CMB measurements are now almost impossible to imagine without the help of cryogenic detectors, direct detection WIMP searches, now led by three cryogenic experiments, together with low-energy particle detection will strongly benefit from the development of these detectors, with their outstanding energy resolution and background identification properties.

%


\begin{thebibliography}{8.}
\addcontentsline{toc}{section}{References}

\bibitem{zwicky} F. Zwicky, {\it Helv. Phys. Acta} {\bf 6} (1933) 110
\bibitem{ramachers} Y. Ramachers, in {\it Proc. XXth Int. Conf. Neutrino Physics and Astrophysics}, May 25-30, 2002, Munich, Germany ; astro-ph/0211500
\bibitem{bergstrom} L. Bergstrom, {\it Rept. Prog. Phys.} {\bf 63} (2000) 793 ; hep-ph/0002126
\bibitem{morales} A. Morales, {\it Nucl. Phys. Proc. Suppl.} {\bf 114} (2003) 39 \bibitem{milgrom} M. Milgrom, {\it Ap. J.} {\bf 270} (1983) 371
and astro-ph/0211446
\bibitem{macho93} C. Alcock et al., (MACHO Collaboration) {\it Nature} {\bf 365} (1993) 621
\bibitem{eros93} E. Aubourg et al., (EROS Collaboration) {\it Nature} {\bf 365} ( 1993) 623
\bibitem{fukuda} Y. Fukuda et al., {\it Phys. Rev. Lett.} {\bf 81} (1998) 1562.
\bibitem{sikivie83} P. Sikivie, {\it Phys. Rev. Lett.} {\bf 51} (1983) 1415 ;
{\it Phys. Rev. Lett.}, {\bf 52} (1984) 695 (Erratum)
\bibitem{combes} D. Pfenniger and F. Combes, {\it Astron. Astroph.} {\bf 285} (1994) 79 ; ibid. 94
\bibitem{scp98} S. Perlmutter et al., {\it Nature} {\bf 391} (1998) 51
\bibitem{stromlo98} A. G. Riess et al., {\it Astrophys. J. } {\bf 116} (1998) 1009
\bibitem{quintessence} I. Zlatev, L. Wang and P. J. Steinhardt, {\it Phys. Rev. Lett.} {\bf 82} (1999) 896
\bibitem{tegmark} M. Tegmark et al., {\it Phys. Rev.} D{\bf 69} (2004) 103501 ; astro-ph/0310723
\bibitem{macho00} C. Alcock et al., (MACHO Collaboration), {\it Ap. J.} {\bf 542} (2000) 281
\bibitem{eros00} P. Tisserand, PhD Thesis, Orsay University (2004), unpublished ; P. Tisserand et al., EROS Collaboration, in preparation
\bibitem{eros03} EROS Collaboration, {\it Astron. Astrophys.} {\bf 400} (2003) 951 ; astro-ph/0212176
\bibitem{ibata98} R. Ibata et al., {\it Ap. J.} {\bf 524} (1998) L1
\bibitem{goldman} B.Goldman, et al. (EROS collaboration), {\it Astron. Astroph.} {\bf 389} (2002) L69
\bibitem{schramm} D. N. Schramm, and M. S. Turner, {\it Rev. Mod. Phys. } {\bf 70} (1998) 303
\bibitem{cyburt} R. H. Cyburt, B. D. Fields, K. A. Olive, {\it Phys.Lett.} B {\bf 567} (2003) 227 ; astro-ph/0302431
\bibitem{cuoco} A. Cuoco, F. Iocco, G. Mangano, G. Miele, O. Pisanti, P.D. Serpico astro-ph/0307213 
\bibitem{hinshaw} G. Hinshaw et al., {\it Ap. J.} {\bf 464} (1996) L17
\bibitem{boomerang} P. de Bernardis et al., {\it Nature} {\bf 404} (2000) 955
\bibitem{maxima} A. T. Lee et al, {\it Ap. J.} {\bf 561} (2001) L1 ; astro-ph/0104459
\bibitem{dasi} J. Kovac, E. M. Leitch, C. Pryke, J. E. Carlstrom, N. W. Halverson, W. L. Holzapfel, {\it Nature} {\bf 420} (2002) 772 ; astro-ph/0209478
\bibitem{archeops} A. Benoit, et al., {\it Astron. Astrophys.} {\bf 399} (2003) L25 ; astro-ph/0210306
\bibitem{cbi} S. Padin, et al., {\it Ap. J.} {\bf 549} (2001) L1
\bibitem{wmap} D.N. Spergel et al., {\it Astrophys. J. Suppl.} {\bf 148} (2003) 97 ; astro-ph/0302209/
\bibitem{initial1} S. P. Ahlen, et al., {\it Phys. Lett.} B {\bf 195} (1987) 603 ; D. O. Caldwell, et al., {\it Phys. Rev. Lett.}, {\bf 61} (1988) 510
\bibitem{initial2} D. O. Caldwell, et al., {\it Phys. Rev. Lett.} {\bf 65} (1990) 1305
\bibitem{wimps} B. W. Lee, S. Weinberg, {\it Phys. Rev. Lett. } {\bf 39} (1977) 165 ; E. Witten, {\it Phys. Rev. } D {\bf 30} (1984) 272 ; K. Greist, {\it Phys. Rev. }, D {\bf 38} (1988) 2357 ; G. Jungman, M. Kamionkowski, K. Griest, {\it Phys. Rep. } {\bf 267} (1996) 195 ; J. Ellis, T. Falk, K. A. Olive, and M. Schmitt, {\it Phys. Lett. } B {\bf 413} (1997) 355 ; J. Edsjo and P. Gondolo, {\it Phys. Rev. } D {\bf 56} (1997) 1879 ; R. Arnowitt and P. Nath, hep-ph/9902237 and {\it Phys. Rev. } D {\bf 60} (1999) 044002, and references therein
\bibitem{caustics} P. Sikivie, {\it Phys.Rev. } D {\bf 60} (1999) 063501 ; {\it Phys. Lett. } B {\bf 432} (1998) 139
\bibitem{drukier84} A.K. Drukier and L. Stodolsky, {\it Phys. Rev. } D {\bf 30} (1984) 2295
\bibitem{goodman} M. W. Goodman and E. Witten, {\it Phys. Rev. } D {\bf 31} (1985) 3059
\bibitem{rich96} J. Rich, {\it Astropart. Phys.} {\bf 4} (1996) 387
\bibitem{annual_modulation} A.K. Drukier, K. Freese and D.N. Spergel, {\it Phys. Rev. } D {\bf 33} (1986) 3495 ; K. Freese, J. Friedman and A. Gould, {\it Phys. Rev. } D {\bf 37} (1988) 3388
\bibitem{directional} D. Spergel, {\it Phys. Rev. } D {\bf 37} (1988) 1353 ; C.J. Copi, J. Heo and L.M. Krauss, {\it Phys. Lett. } B {\bf 461} (1999) 43
\bibitem{sun_capture} J. Silk, K. Olive and M. Srednicki, {\it Phys. Rev. Lett. } {\bf 55} (1985) 257 ; L.M. Krauss, M. Srednicki and F. Wilczek, {\it Phys. Rev. } D {\bf 33} (1986) 2079
\bibitem{indirect} J. Silk, K. Olive and M. Srednicki, {\it Phys. Rev. Lett.} {\bf 55}, 257 (1985); L.M. Krauss, M. Srednicki and F. Wilczek, {\it Phys. Rev.} {\bf D33}, 2079 (1986)
\bibitem{gondolo_silk} P. Gondolo, J. Silk, {\it Phys. Rev. Lett.} {\bf 83} (1999) 1719
\bibitem{heidelberg94} M. Beck. et al., Heidelberg-Moscow Collaboration, {\it Phys. Lett. } B {\bf 336} (1994) 141
\bibitem{heidelberg99} L. Baudis et al., {\it Phys. Rev.} {\bf D59}, 022001 (1999)
\bibitem{hdms01} L. Baudis, et al., {\it Phys.Rev.} {\bf D63}, 022001 (2001)
\bibitem{igex} A. Morales et al., {\it Phys. Lett.} {\bf B532}, 8 (2002), hep- ex/0110061
\bibitem{dama96} R. Bernabei et al., {\it Phys. Lett.} {\bf B389}, 757 (1996)
\bibitem{ukdmc96} P. F. Smith et al., {\it Phys. Lett.} {\bf B379}, 299 (1996)
\bibitem{gerbier99} G. Gerbier et al., {\it Astropart. Physics } {\bf 11}, 287 (1999)
\bibitem{dama9899} R. Bernabei et al., {\it Phys. Lett.} {\bf B424}, 195 (1998) ; {\it Phys. Lett.} {\bf B450}, 448 (1999)
\bibitem{ejiri} H. Ejiri et al., in "Proceedings of the Second International Workshop on the Identification of Dark Matter", edited by N. J. C. Spooner and V. Kudryavtsev (World Scientific, Singapore, 1999), p. 323
\bibitem{dama00} R. Bernabei et al., {\it Phys. Lett.} {\bf B480}, 23 (2000)
\bibitem{dama03} R. Bernabei et al., {\it Riv. Nuovo Cim.} {\bf 26} (2003) 1 ; astro-ph/0307403
\bibitem{lewin_smith} J.D. Lewin and P.F. Smith, {\it Astropart. Phys.} 6, 87 (1996)
\bibitem{edelw01} A. Benoit et al., {\it Phys.Lett.} {\bf B513}, 15 (2001) and astro-ph/0106094/
\bibitem{edelw02} A. Benoit et al., {\it Phys. Lett.} {\bf B545}, 43 (2002)
\bibitem{edelw03} V. Sanglard, the EDELWEISS collaboration, astro-ph/0406537, to appear in {\it Proceedings of Rencontres de Moriond, "Exploring the Universe"}, La Thuile, Italy (March 28 - April 4, 2004)
\bibitem{copi} C.J. Copi, L.M. Krauss, {\it Phys.Rev.} {\bf D67} (2003) 103507 ; astro-ph/0208010/
\bibitem{kurylov} A. Kurylov and M. Kamionkowski, {\it Phys. Rev.} D{\bf 69} (2004) 063503 ; hep-ph/0307185
\bibitem{savage} C. Savage,  P. Gondolo,  K. Freese, astro-ph/0408346, submitted to {\it Phys. Rev.}~D
\bibitem{zeplin} N. J. T. Smith et al., in {\it Proceedings of the 4th International Workshop on Identification of Dark Matter (idm2002)}, eds. N.J. Spooner and V. Kudryavtsev, (World Scientific, Singapore, 2003)
\bibitem{dmtools} R. Gaitskell and V. Mandic, Interactive WIMP Limits Plotter, http://dmtools.berkeley.edu/limitplots/
\bibitem{shutt92} T. Shutt, et al., {\it Phys. Rev. Lett.} {\bf 69}, 3452 (1992) ; {\it Phys. Rev. Lett.} {\bf 69}, 3531 (1992)
\bibitem{lhote} D. L'H\^{o}te, X-F. Navick, R. Tourbot, J. Mangin and F. Pesty, {\it J. Appl. Phys.} {\bf 87} (2000) 1507
\bibitem{bravin} M. Bravin et al., {\it Astropart. Phys.} {\bf 12}, 107 (1999)
\bibitem{meunier} P. Meunier et al., {\it Appl. Phys. Lett.} {\bf 75} (1999) 1335
\bibitem{coron} N. Coron, et al., {\it Astron. Astrophys.} {\bf 278} (1993) L31
\bibitem{cebrian01} S.Cebrian, et al., {\it Astropart.Phys.} {\bf 15} (2001) 79 ; astro-ph/0004292
\bibitem{cdms00} R. Abusaidi et al., {\it Phys. Rev. Lett.} {\bf 84}, 5699 (2000)
\bibitem{cdms02} D. Abrams et al., CDMS collaboration, {\it Phys.Rev.} {\bf D66}, 122003 (2002), astro-ph/0203500
\bibitem{alessandrello96} A. Alessandrello, et al., {\it Nucl. Instrum. Meth.} A {\bf 370} (1996) 241
\bibitem{alessandrello97} A. Alessandrello, {\it Phys. Lett.} B {\bf 408} (1997) 465
\bibitem{ootani} W. Ootani et al., {\it Phys. Lett.} B {\bf 461} (1999) 371; {\it Nucl. Instr. Meth.}, A {\bf 436} (1999) 233
\bibitem{stefano} P. Di Stefano et al., {\it Astropart. Phys.} {\bf 14}, 329 (2001) ; astro-ph/0004308/ 
\bibitem{young} B. Young, et al., {\it Nucl. Instrum. and Meth.} A{\bf 311}
(1992) 195.
\bibitem{irwin_review} K. D. Irwin et al., {\it Rev. Sci. Instr.} {\bf 66} (1995) 5322
\bibitem{cabrera} B. Cabrera et al., {\it Appl. Phys. Lett.} {\bf 73} (1998)
735.
\bibitem{lindhard} J.~Lindhard et al., {\it Mat. Fys. Medd. Dan. Vid. Selsk} {\bf 10} (1963) 1
\bibitem{spooner91} N.J.C Spooner, et al., {\it Phys. Lett.} B {\bf 273} (1991) 333
\bibitem{clarke} R. M. Clarke et al., {\it Appl. Phys. Lett.} {\bf 76} (2000) 2958
\bibitem{navick} X.F.~Navick {\it et al.}, {\it Nucl. Instrum. Meth.} A {\bf 444} (2000) 361
\bibitem{cdms03} CDMS Collaboration, hep-ex/0306001, {\it Phys. Rev.} D{\bf 68} (2003) 082002 
\bibitem{cdms04} D. Akerib et al., the CDMS Collaboration, {\it Phys. Rev. Lett} {\bf 93} (2004) 211301 ; astro-ph/0405033
\bibitem{mirabolfathi} N. Mirabolfathi, et al., {\it AIP Conf. Proc.} {\bf 605} (2002) 517 
\bibitem{sicane} E. Simon, et al., {\it Nucl. Instrum. Meth.} A {\bf 507} (2003) 643 ; astro-ph/0212491
\bibitem{ramo} S. Ramo, in {\it Proceedings of the I.R.E.} {\bf 27} (1939) 584
\bibitem{llacer} J. Llacer, E.E. Haller, R.C. Cordi, {\it IEEE Trans. Nucl. Sci.}, {\bf 24} (1977) 53
\bibitem{luke92} P.N. Luke, C.P. Cork, N.W. Madden, C.S Rossington, M.F. Wesela, {\it IEEE Trans. Nucl. Sci.} {\bf 39} (1992) 590
\bibitem{luke94} P.N. Luke, C.S Rossington, M.F. Wesela, {\it IEEE Trans. Nucl. Sci.} {\bf 41} (1994) 1074
\bibitem{shutt00} T. Shutt et al., {\it Nucl. Instr. Meth.} A {\bf 444} (2000)340
\bibitem{shutt01} T. Shutt et al., in {\it Proc. 9th Int. Workshop on Low Temperature Detectors}, AIP conference proceedings {\bf 605} (2001) 513
\bibitem{neganov} B. Neganov and V.~Trofimov, USSR patent No 1037771, {\it Otkrytia i izobreteniya} {\bf 146} (1985) 215
\bibitem{luke} P.N.~Luke, {\it J. Appl. Phys.} {\bf 64} (1988) 6858
\bibitem{booth} N. Booth, {\it Appl. Phys. Lett.} {\bf 50}, 293 (1987)
\bibitem{irwin} K. D. Irwin et al., {\it Appl. Phys. Lett.} {\bf 66} (1995) 1998
\bibitem{welty} R.P. Welty and J.M. Martinis, {\it IEEE Trans. Appl. Superconduct.} {\bf 3} (1993) 2605
\bibitem{hellmig} J. Hellmig et al., {\it Nucl. Instrum. Meth.} A {\bf 444} (2000) 308
\bibitem{penn} M.J. Penn, B.L. Dougherty, B. Cabrera, R.M. Clarke, {\it J. Appl. Phys.} {\bf 79} (1996) 8179
\bibitem{broniato03} A. Broniatowski, et al., in {\it Proceedings of the 10th International Workshop on Low temperature Detectors}, Genoa, 7-11 July 2003, {\it Nucl. Instr. Meth.} A {\bf 520} (2004) 178
\bibitem{censier03} B. Censier, et al., in {\it Proceedings of the 10th International Workshop on Low temperature Detectors}, Genoa, 7-11 July 2003, {\it Nucl. Instr. Meth.} A {\bf 520} (2004) 156
\bibitem{broniato01} A. Broniatowski, et al., in {\it Low Temperature Detectors}, eds. F.S. Porter et al., {\it AIP Conference Proceedings} {\bf 605}, 521 (2001)
\bibitem{chardin03} G. Chardin, et al., in {\it Proceedings of the 10th International Workshop on Low temperature Detectors}, Genoa, 7-11 July 2003, {\it Nucl. Instr. Meth.} A {\bf 520} (2004) 145
\bibitem{kudryavstev} V. A. Kudryavtsev et al., {\it Phys. Lett.} B {\bf 452} (1999) 167
\bibitem{bobin} C. Bobin et al., {\it Nucl. Instr. Meth.} A {\bf 386} (1997) 453
\bibitem{fiorini_scint} A. Alessandrello et al., {\it Phys. Lett.} B {\bf 420} (1998) 109
\bibitem{cebrian03} S. Cebrian, et al., {\it Phys. Lett.} B {\bf 563} (2003) 48
\bibitem{tecnomusiq} http://www.e15.physik.tu-muenchen.de/Tecnomusiq/Tecnomusiq.html 
\bibitem{weber} M. F. Weber, C. Stover, L. Gilbert, T. Nevitt, and A. Ouderkirk, {\it Science} {\bf 287} (2000) 2451
\bibitem{angloher04} G. Angloher, et al., CRESST collaboration, astro-ph/0408006
\bibitem{bradley} D.I. Bradley et al., {\it Phys. Rev. Lett.} {\bf 75} (1995) 1887
\bibitem{santos} D. Santos, in {\it Sources and detection of Dark Matter and Dark Energy in the Universe }, ed. D. B. Cline, (Springer, Berlin, 2001)
\bibitem{bernas} H. Bernas, et al., {\it Phys. Lett.} A {\bf 24} (1967) 721
\bibitem{orpheus} M. Abplanalp et al., {\it Nucl. Instrum. and Meth.} A{\bf 370} (1996) 227
\bibitem{orpheus_idm02} K. Borer, et al., in {\it Proceedings of the 4th International Workshop on the  Identification of Dark Matter} (IDM 02), York, England, eds. N.J.C. Spooner and V. Kudryavtsev, (World Scientific, Singapore, 2003), pp. 332-337
\bibitem{miuchi} K.Miuchi, et al., {\it Astropart. Phys.} {\bf 19} (2003) 135 ; astro-ph/0204411
\bibitem{ellis_flores} J. Ellis and R.A. Flores, {\it Phys. Lett.} B {\bf 263} (1991) 259
\bibitem{superk} A. Habig, for the Super-Kamiokande Collaboration, hep-ex/0106024
\bibitem{amanda} E. Andres et al., {\it Astropart. Phys.} {\bf 13}, 1 (2000)
\bibitem{kamionkowski} M. Kamionkowski, K. Griest, G. Jungman, and B. Sadoulet, {\it Phys. Rev. Lett.} {\bf 74} (1995) 5174
\bibitem{marcillac03} P. de Marcillac, et al., in {\it Proceedings of the 10th International Workshop on Low temperature Detectors}, Genoa, 7-11 July 2003, {\it Nucl. Instr. Meth.} A {\bf 520} (2004) 159
\bibitem{rich88} J. Rich and M. Spiro, {\it Saclay report} DPhPE-88-04 (1988)
\bibitem{buckland} K.N. Buckland, M.J. Lehner, G.E. Masek, and M. Mojaver, {\it Phys. Rev. Lett.} {\bf 73} (1994) 1067
\bibitem{lanou} R. E. Lanou, Nucl. {\it Phys. B Proc. Suppl.} {\bf 77} (1999) 55
\bibitem{bandler} S.R. Bandler et al., {\it Phys. Rev. Lett.} {\bf 74} (1995) 3169
\bibitem{collar} J.I. Collar and F.T. Avignone III, {\it Nucl.
Instr. Methods} B {\bf 95} (1995) 349
\bibitem{lehner} M.J. Lehner, K.N. Buckland and G.E. Masek,
{\it Astropart. Phys.} {\bf 8} (1997) 43
\bibitem{snowden_ifft95} D.P. Snowden-Ifft, E.S. Freeman and P.B. Price, {\it Phys. Rev. Lett.} {\bf 74} (1995) 4133
\bibitem{snowden_ifft97} D. P. Snowden-Ifft and A. J. Westphal, {\it Phys. Rev. Lett.} {\bf 78} (1997) 1628
\bibitem{drift00} D. P. Snowden-Ifft, C. J. Martoff, J. M. Burwell, {\it Phys.Rev.} D {\bf 61} (2000) 101301
\bibitem{cresst04} G. Angloher et al., the CRESST Collaboration, astro-ph/0408006, submitted to {\it Astropart. Phys.}
\bibitem{sanglard04} V. Sanglard et al., the EDELWEISS Collaboration, astro-ph/0406537, in {\it Proceedings of the Rencontres de  Moriond - Cosmology : Exploring the Universe}, (March 28 - April 4, 2004), La Thuile (Italy)
\bibitem{dama_oneton} R. Bernabei et al., {\it Astropart. Phys.} {\bf 4} (1995) 45
\bibitem{genius} H.V. Klapdor-Kleingrothaus, {\it Nucl. Phys.} {\bf B110}, 364 (2002), and hep-ph/ 0206249
\bibitem{genius_tf} L. Baudis et al., {\it Nucl. Instrum. Meth.} {\bf A481} (2002) 149, hep-ex/0012022
\bibitem{majorana} L. De Braeckeleer, for the MAJORANA Collaboration, in {\it Proceedings of the Carolina Symposium on Neutrino Physics}, South Carolina, 10-12 Mar 2000, pp. 325-339.
\bibitem{cuoricino} S.Pirro et al. {\it Nucl. Instr. Meth.} A {\bf 444} (2000) 71
\bibitem{cuore} C. Arnaboldi et al., CUORE Collaboration, {\it Astropart. Phys.} {\bf 20} (2003) 91 and hep-ex/0302021
\bibitem{ramachers_annual} Y. Ramachers, M. Hirsch, H.V. Klapdor-Kleingrothaus, Eur. {\it Phys. J.} A {\bf 3} (1998) 93
\bibitem{hasenbalg} F. Hasenbalg, {\it Astropart. Phys.} {\bf 9} (1998) 339
\bibitem{nautilus} P. Astone et al., {\it Astropart. Phys.} {\bf 7} (1997) 231
\bibitem{grail} G. Frossati, {\it J. Low Temp. Phys.} {\bf 101} (1995) 81
\bibitem{cryo_review2} H. Kraus, {\it Supercond. Sci. Technol.} {\bf 9} (1996) 827
\bibitem{marnieros} S. Marnieros, L. Berge, A. Juillard, L. Dumoulin, {\it Phys. Rev. Lett.} {\bf 84} (2000) 2469 ; A. Juillard, PhD Thesis, Orsay, (1999), unpublished
\bibitem{borexino} F. Gatti, G. Morelli, G. Testera, S. Vitale, {\it Nucl. Instrum. Meth.} A {\bf 370} (1996) 609 
\bibitem{karmen} B. Zeitnitz, G. Drexlin, {\it Interdisciplinary Science Reviews} (ISR) {\bf 18} (1993) 313
\bibitem{ellis00} J. Ellis, A. Ferstl, K. A. Olive, {\it Phys.Lett.} {\bf B481}, 304 (2000)



\end{thebibliography}
\end{document}